\documentclass[english,11pt]{article}

\usepackage{subfigure}
\usepackage{graphicx}
\usepackage[colorlinks=true,citecolor=blue,urlcolor=blue]{hyperref}

\usepackage[latin9]{inputenc}
\usepackage{float}
\usepackage{bm}

\usepackage{amsmath}
\usepackage{amsthm,amssymb,amsfonts}
\usepackage{upgreek}
\usepackage[hmargin=0.9in]{geometry}
\usepackage[textsize=footnotesize]{todonotes}


\usepackage{babel}

\newcommand{\order}{{\mathcal O}}
\newcommand{\Dx}{\Delta x}
\newcommand{\Dy}{\Delta y}
\newcommand{\bx}{{\bf x}}
\def\ceff{c_\text{eff}}

\begin{document}

\title{Diffractons: solitary waves created by diffraction in periodic media}
\author{David I. Ketcheson\thanks{King Abdullah University of Science and Technology (KAUST), CEMSE Division. E-mail: 
		\url{david.ketcheson@kaust.edu.sa}} \and
	Manuel Quezada de Luna\thanks{Department of Mathematics, Texas A\&M University. College Station, Texas
		77843, USA. E-mail: \url{mquezada@math.tamu.edu}}}

\maketitle

\abstract{
A new class of solitary waves arises in the solution of nonlinear
wave equations with constant impedance and no dispersive terms.
They depend on a balance between nonlinearity and a dispersion-like effect due
to spatial variation in the sound speed of the medium.
A high-order homogenized model confirms this effective dispersive behavior and
its solutions agree well with those obtained by direct simulation of the variable-coefficient
system.  These waves are observed to be long-time stable, globally attracting
solutions that arise in general as solutions to nonlinear wave problems with
periodically-varying sound speed.  They share some properties with known classes
of solitary waves, but possess important differences as well.
}

\section{Introduction}
Many nonlinear wave equations are known to have solitary wave (or soliton)
solutions.  These partial differential equations, such as the Korteweg-de Vries
or nonlinear Schrodinger equation, include both nonlinear and dispersive terms.
Solitary wave solutions arise through a balance between nonlinear and dispersive
effects.

Solitary wave solutions have also been observed in simulations of
one-dimensional periodic elastic media 
with spatially-varying impedance and no explicit dispersion.  Such media
exhibit an effective dispersion -- the result of reflections due to the material structure
\cite{leveque2003,ketchesonphdthesis,Ketcheson_LeVeque_2011,quezada2013}.
This effective dispersion is present only in media with spatially-varying impedance:
waves in one-dimensional media with uniform impedance 
behave -- at the macroscopic scale -- essentially like waves
in a homogeneous medium \cite{santosa1991,leveque2003,Ketcheson_LeVeque_2011}.

\begin{figure}
\begin{centering}
\includegraphics[scale=0.40]{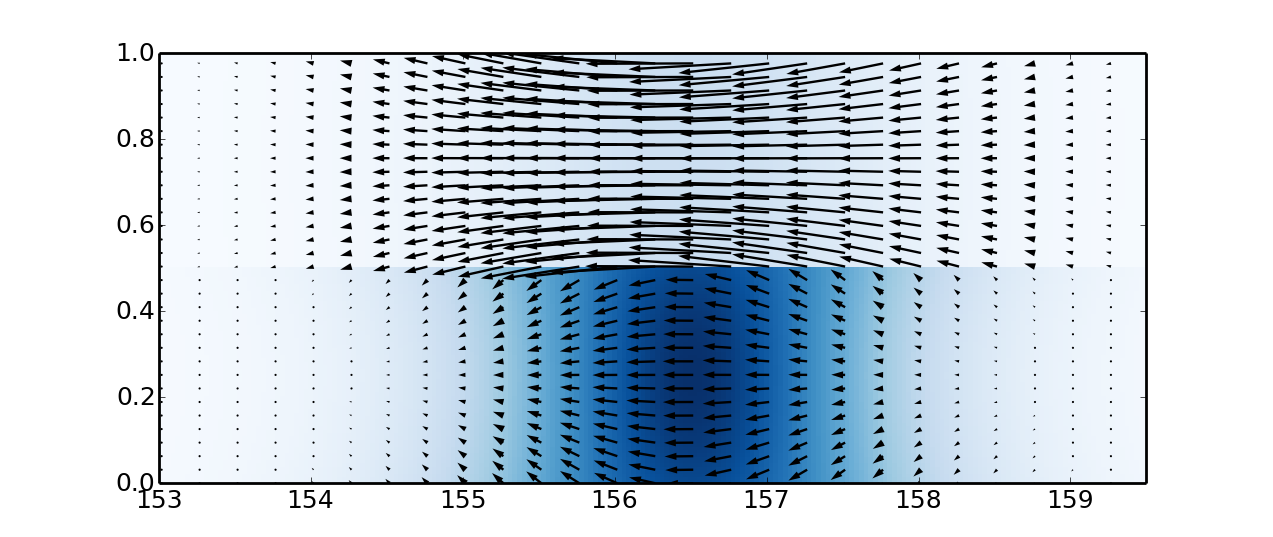}
\par\end{centering}
\caption{Structure of a diffracton.  Strain is shown in blue, and vectors represent
the material velocity.  The solitary wave travels only to the right, but diffraction is
evident in the (vertical) material velocities.\label{fig:diffracton}}
\end{figure}

Herein we report the discovery of solitary wave solutions to a first-order 
hyperbolic system with {\em no dispersive terms} and {\em no reflection}.
These solitary waves arise in two-dimensional periodic media in which {\em the
impedance is constant} and only the sound speed varies.
The mechanism responsible for the effective dispersion that leads to these waves
is diffraction~\cite{QK2013}.
We thus refer to these wave as {\em diffractons}; a typical diffracton
is shown in Figure \ref{fig:diffracton}.  Notice that, although this solitary
wave simply translates to the right, vertical velocities appear in the solution
due to diffraction.
Computational evidence suggests that diffractons are globally
attracting solutions to quite general classes of nonlinear wave equations in
periodic materials.

In this work, we investigate diffractons through computation and analysis.  
The paper is organized as follows.  
In Section \ref{sec:diffractons}, we present the model, materials, and
waves that are the subject of this study.
In Section \ref{sec:homog} we derive an effective model for 2D
nonlinear waves in periodic media based on homogenization. 
In Section \ref{sec:properties} we study symmetries and interactions of diffractons.
Finally, in Section \ref{sec:generalizations},
we show that diffractons arise in a wide range of settings.

 
\section{Solitary waves in non-dispersive, constant-impedance, periodic media\label{sec:diffractons}}
We are interested in the behavior of multidimensional waves in nonlinear,
spatially-varying media.  Essentially the simplest model of this kind is
\begin{align}\label{p-system}
  \epsilon_{tt}-\nabla\cdot\left(\frac{1}{\rho(\bx)} \nabla\sigma(\epsilon,\bx)\right) & = 0,
\end{align}
which may be viewed as a multi-dimensional analog of the $p$-system.
We use the notation of elasticity, for consistency with related 
work \cite{leveque2003,quezada2013};
thus $\epsilon$ is the strain, $\rho$ is the density, and $\sigma$ is the stress.
If the stress-strain function is linear, i.e.
$ \sigma(\epsilon,\bx) = K(\bx) \epsilon, $
then \eqref{p-system} is just the variable-coefficient linear wave equation.

If the stress-strain relation is nonlinear, then
solutions of \eqref{p-system} often involve shock singularities.
In most of what follows, we take
  \begin{align} \label{nonlinear_stress_relation}
    \sigma(\epsilon,\bx) & = \exp(K(\bx)\epsilon)-1.
  \end{align}
Here $K(\bx)$ is referred to as the bulk modulus. 
The particular relation \eqref{nonlinear_stress_relation} is convenient for
performing homogenization, but the phenomenon under study seems to appear
when $\sigma$ is any nonlinear function.

Solutions of \eqref{p-system} with the stress relation 
\eqref{nonlinear_stress_relation} often involve shock singularities.
In order to determine entropy-satisfying weak
solutions, we write \eqref{p-system} as a first-order
hyperbolic system of conservation laws:
\begin{subequations} \label{2D psystem conservation form}
  \begin{equation}
    {\bf q}_{t}+{\bf f}({\bf q}, \bx)_x+{\bf g}({\bf q}, \bx)_y={\bf 0},
  \end{equation}
where
  \begin{align}
    {\bf q} & = \begin{bmatrix} \epsilon\\ \rho(\bx) u \\ \rho(\bx)  v \end{bmatrix},
    & {\bf f}({\bf q}, {\bf x}) & =\begin{bmatrix} -u \\ -\sigma(\epsilon,{\bf x})\\ 0\end{bmatrix},
    & {\bf g}({\bf q}, {\bf x}) & =\begin{bmatrix} -v\\ 0\\ -\sigma(\epsilon,{\bf x})\end{bmatrix}.
  \end{align}
\end{subequations}
Here $u$ and $v$ are the $x$- and $y$-components of velocity,
${\bf q}$ is the vector of conserved quantities, and ${\bf f}, {\bf g}$
are the components of the flux in the $x$- and $y$-directions, respectively.

We consider media in which the material parameters are uniform in one direction
($x$) and vary periodically in the other ($y$) with unit period; i.e. $K =
K(y)$, $\rho=\rho(y)$ with
\begin{align*}
K(y+1) & = K(y), & \rho(y+1) & = \rho(y).
\end{align*}
The linearized sound speed $c(y) = \sqrt{K/\rho}$ and linearized impedance
$Z(y) = \sqrt{K \rho}$ are thus also periodic and will play a central role
in our analysis.

\begin{figure}
\begin{centering}
  \includegraphics[scale=0.35]{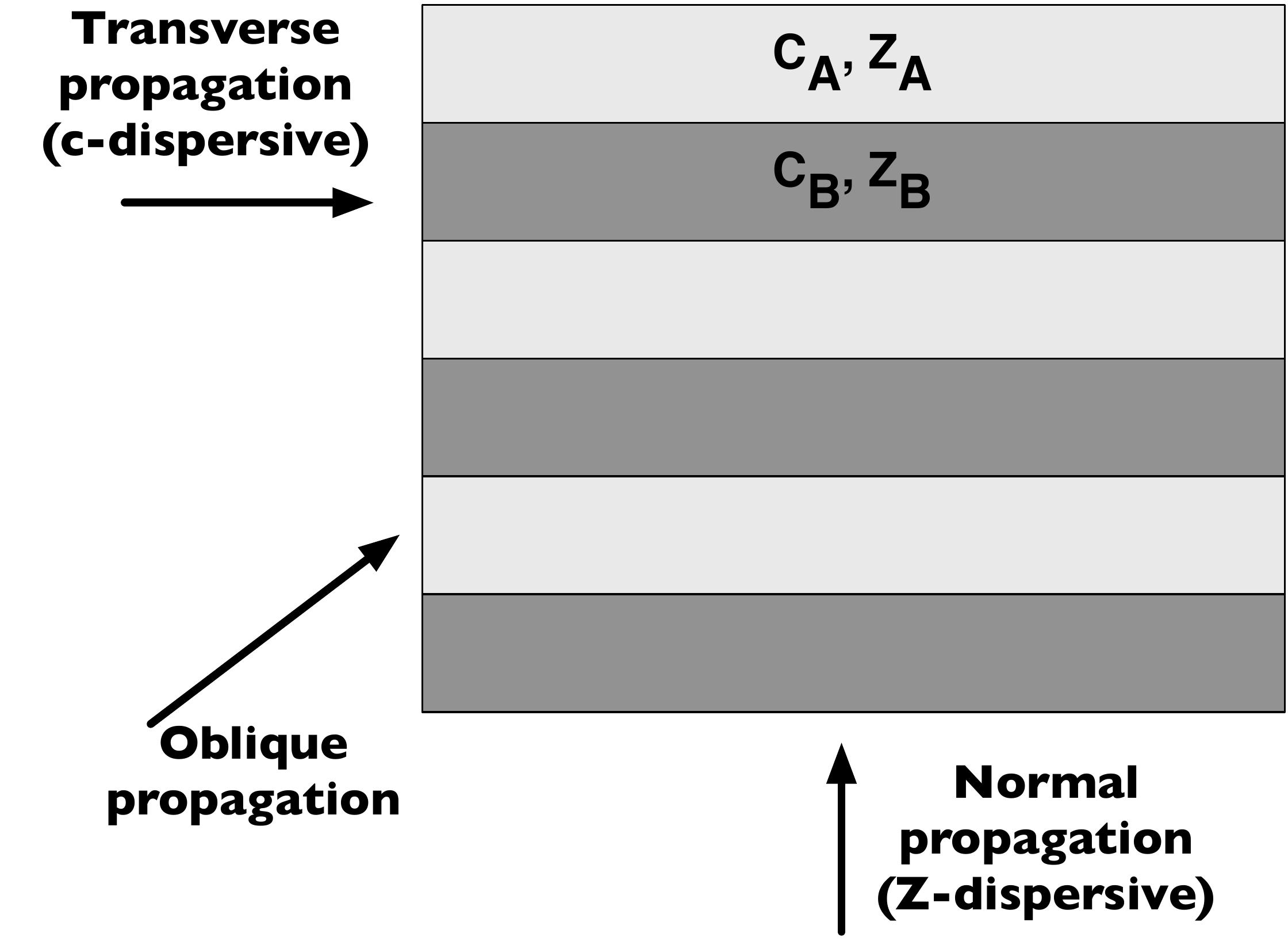}
\par\end{centering}
\caption{2D wave propagation in a one-dimensionally-periodic medium.
A piecewise-constant layered medium is shown.\label{fig:layered}}
\end{figure}

We primarily investigate piecewise-constant (layered) media, as
shown in Figure \ref{fig:layered}:
\begin{align} \label{discontinuous medium}
 K(y),\rho(y) & = \begin{cases}
                    (K_A,\rho_A) \mbox{ if } \left(y-\lfloor y\rfloor-\frac{1}{2}\right)<0 \\
                    (K_B,\rho_B) \mbox{ if } \left(y-\lfloor y\rfloor-\frac{1}{2}\right)>0.
  \end{cases}
\end{align}
We use the terms {\em normal wave} and {\em transverse wave} to indicate
plane wave perturbations traveling orthogonal to or parallel to the layer
interfaces, as indicated in Figure \ref{fig:layered}.
Propagation of a normally-incident plane wave can (by symmetry) be modelled as a one-dimensional
problem; in this case our model reduces to that studied in \cite{leveque2003}.
There it was observed that solitary waves can form when the impedance constrast is
sufficiently high, due to the net effect of reflections.
When the impedance is uniform, normal waves behave similarly to solutions of
Burgers' equation, leading to shock formation and $N$-wave decay~\cite{Ketcheson_LeVeque_2011}.

In the present work, we are mainly interested in 
transversely propagating waves, i.e. those arising from perturbations that are
uniform in $y$.  As we will see, such perturbations can 
lead to solitary wave formation even when the impedance is constant.
The effect responsible for this is diffraction, which appears whenever the
medium sound speed varies.  Note that, unlike normal perturbations, transverse
perturbations represent a genuinely two-dimensional phenomenon.

Let us conduct a few computational experiments to indicate the types of
behavior possible for transversely propagating waves.
We consider a wave entering the domain from the left generated by a moving wall
boundary condition:
\begin{align} \label{moving_wall}
 \rho(y) u(0,y,t) & = \begin{cases}
                      m(1+\cos{(\pi(t-10)/10)}) & 0 \le t \le 20, \\
                      0						& t \ge 20.
  \end{cases}
\end{align}
Here we take the peak momentum as $m=0.1$.
By symmetry, the problem can be solved by considering a single period of the
medium and periodic boundary conditions in $y$.  We compute solutions to
\eqref{p-system} using the finite volume solver PyClaw \cite{pyclaw-sisc,clawpack}
with the Riemann solvers described in \cite{quezada2013}.
We consider the solution after the perturbation has travelled a distance of more
than 300 material periods.

\begin{figure}
 \begin{centering}
  \subfigure[Linear medium, constant $c$]{
  		\includegraphics[scale=0.3]{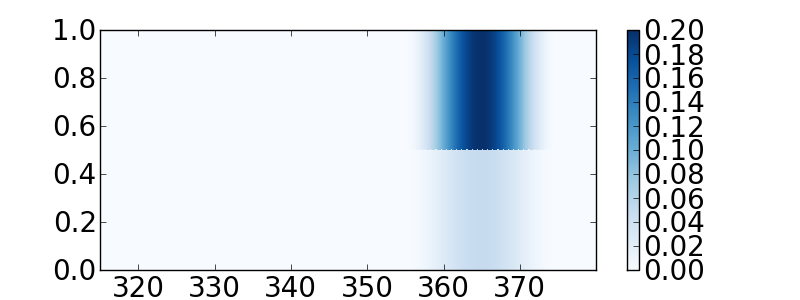}
		\includegraphics[scale=0.3]{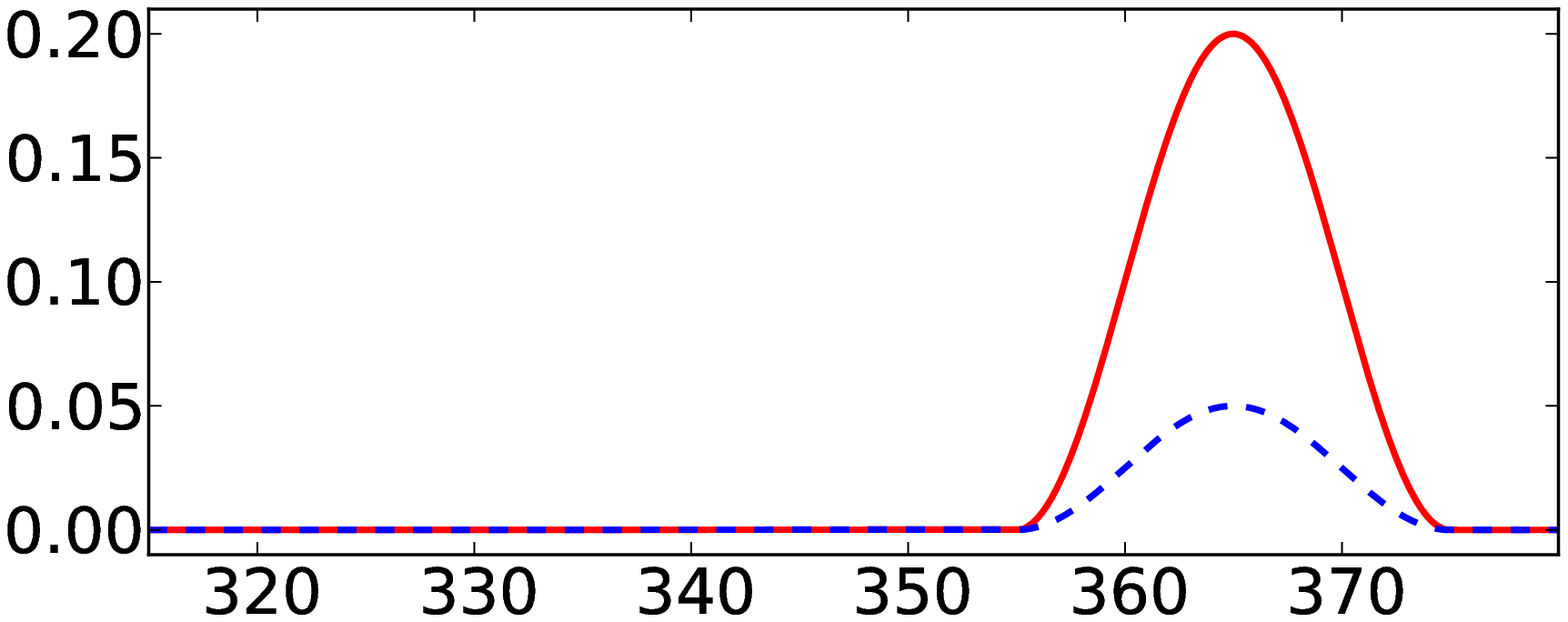}
		\label{fig:linear_constant}}
  \subfigure[Linear medium, variable $c$]{
  		\includegraphics[scale=0.3]{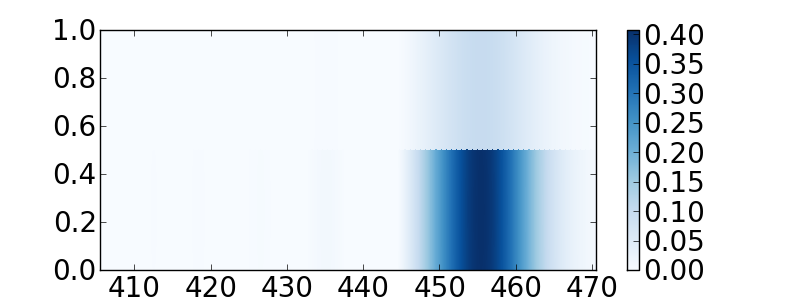}
		\includegraphics[scale=0.3]{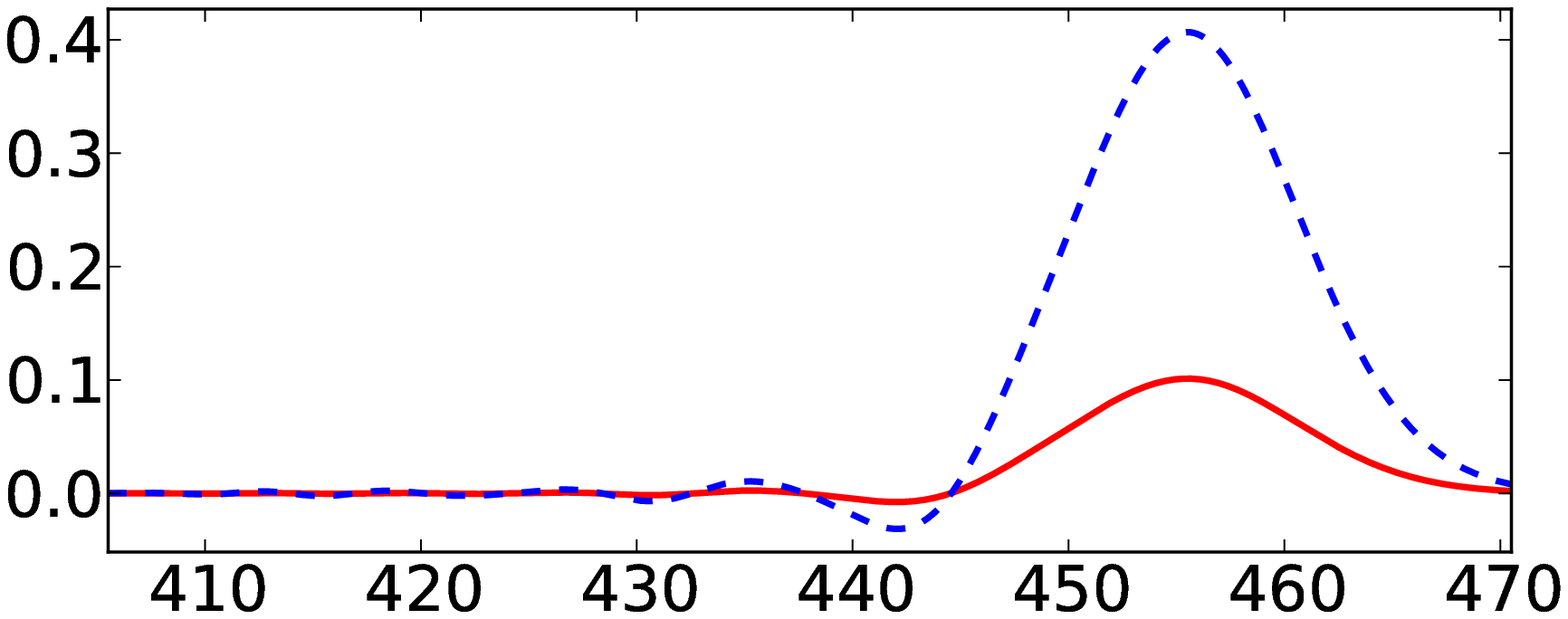}
		\label{fig:linear_variable}}
  \subfigure[Nonlinear medium, constant $c$]{
  		\includegraphics[scale=0.3]{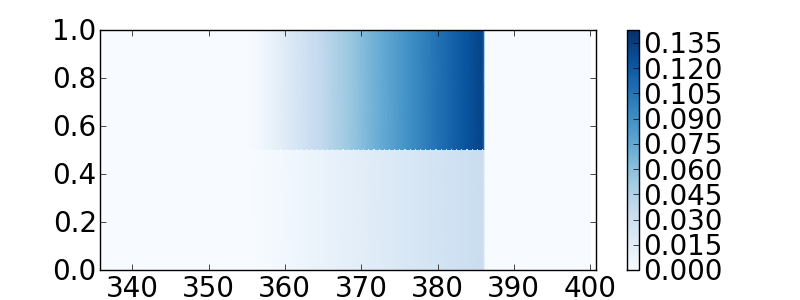}
		\includegraphics[scale=0.3]{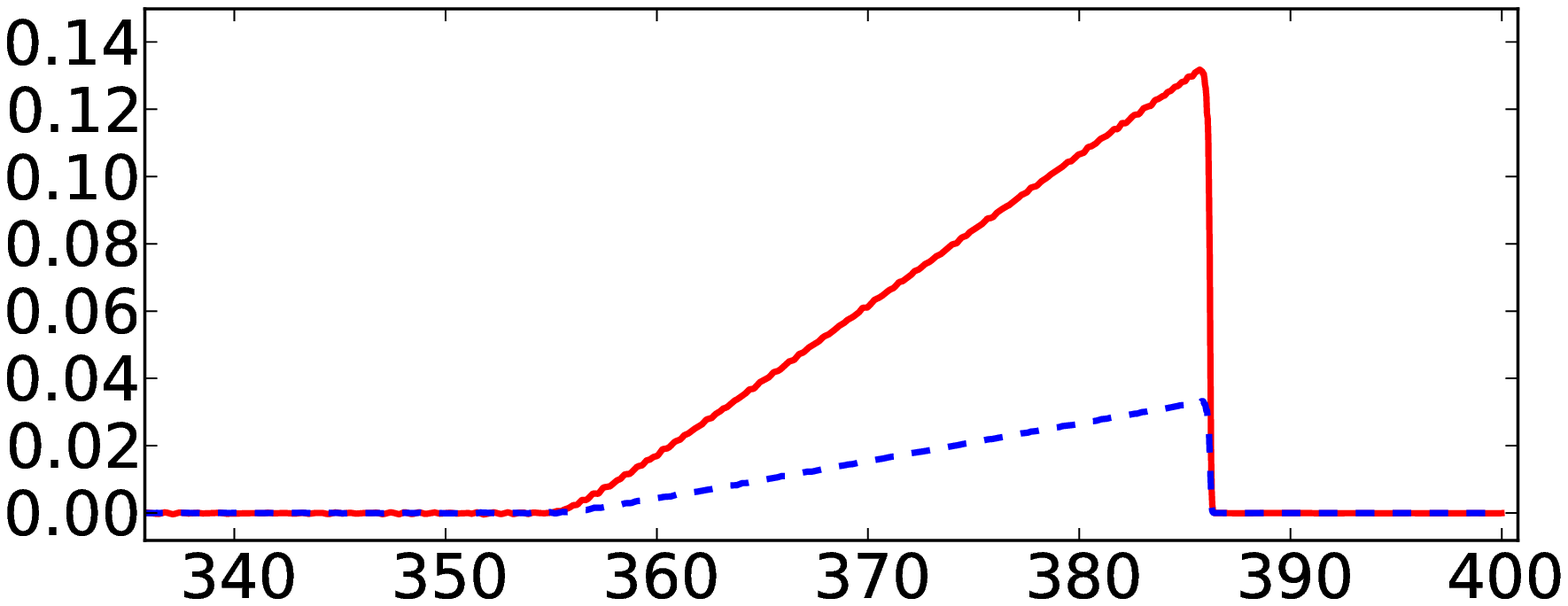}
		\label{fig:nonlinear_constant}}
  \subfigure[Nonlinear medium, variable $c$]{
  		\includegraphics[scale=0.3]{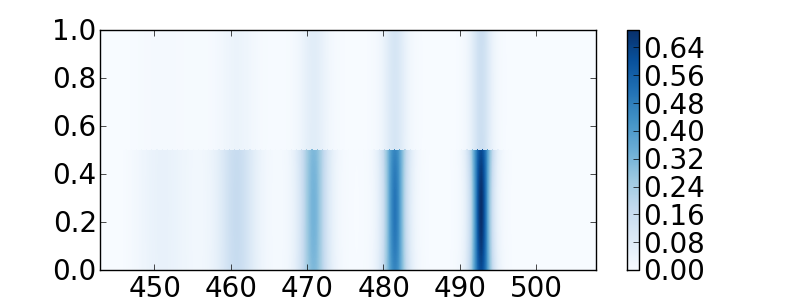}
		\includegraphics[scale=0.3]{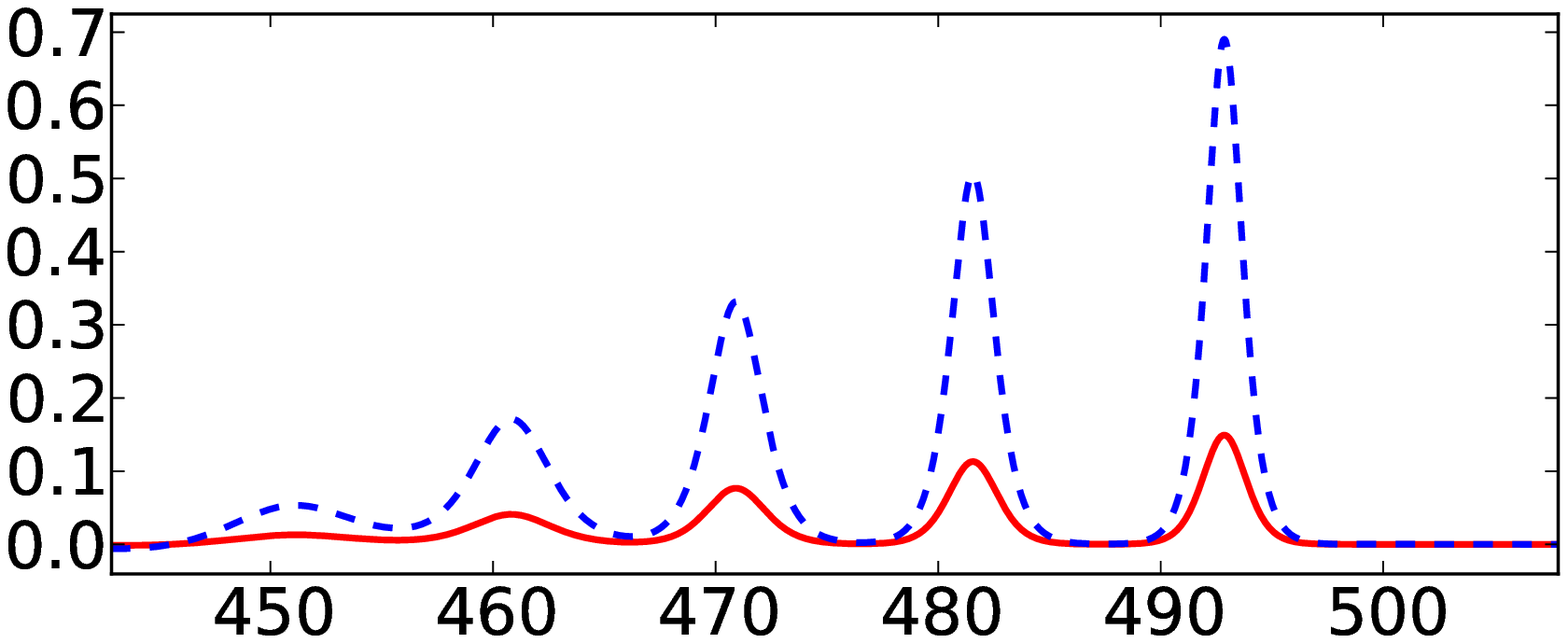}
		\label{fig:nonlinear_variable}}
\par\end{centering}
\caption{Strain at $t=375$ for four experiments.  We show surface plots (left, note high aspect ratio)
and $y$-slices (right) at the middle of material A (blue) and material B (red).  
\label{fig:intro_experiments}}
\end{figure}

First we consider the simpler case of linear wave propagation, by taking
the linear constitutive relation
\begin{align*}
\sigma(\epsilon,\bx) = K(y) \epsilon
\end{align*}
in place of \eqref{nonlinear_stress_relation}.
It turns out that the resulting solution depends qualitatively on
whether the sound speeds in the two materials are equal or not.
Figure \ref{fig:linear_constant} (note the very high aspect ratio)
shows the solution obtained in a medium with parameters
\begin{subequations} \label{cmatched}
\begin{align}
c_A & = 1, & Z_A & = 4, \\
c_B & = 1, & Z_B & = 1,
\end{align}
\end{subequations}
obtained by taking $K_A=\rho_A=4$ and $K_B=\rho_B=1$.
Because $c_A=c_B$, the initial perturbation travels at constant velocity
without changing shape.

Figure \ref{fig:linear_variable}
shows a solution obtained with 
\begin{subequations} \label{zmatched}
\begin{align}
c_A & = 5/8, & Z_A & = 1, \\
c_B & = 5/2, & Z_B & = 1,
\end{align}
\end{subequations}
obtained by taking $K_A = 1/\rho_A = 5/8$ and
$K_B = 1/\rho_B = 5/2$.
Since $c_A \ne c_B$
the initial perturbation undergoes diffraction, leading to an
effective dispersion.  High frequencies travel more slowly, so
the solution develops an oscillatory tail.  We emphasize that
this {\em effective dispersion} is a macroscopic effect
of the material microstructure; clearly, no dispersive terms 
appear in the model equations.  This effect has been studied
in detail for linear waves in \cite{QK2013}.

Next we consider the same two scenarios, but with the nonlinear
stress relation \eqref{nonlinear_stress_relation}.
Figure \ref{fig:nonlinear_constant} shows the solution for
a medium with $c_A=c_B=1$.  The solution behaves like a one-dimensional
perturbation in a homogeneous medium.  A shock forms and leads
to gradual decay of the solution amplitude.

Finally, in Figure \ref{fig:nonlinear_variable}, we consider the
main case of interest: a nonlinear medium with $c_A\ne c_B$.
The combination of nonlinearity and effective (diffractive) dispersion leads
to the formation of a train of solitary waves.  We refer to these
waves as {\em diffractons}.

\section{Homogenization\label{sec:homog}}
Analysis of the wave equation \eqref{p-system} is complicated by the
presence of variable coefficients.  Here we give a homogenized approximation
(with constant coefficients), and use it to investigate diffractons.

In \cite{QK2013}, high order homogenized equations are derived for {\em linear} acoustic waves
in a medium of the type considered in this work.
It is assumed that $\lambda$, the typical wavelength of the solution,
is large compared to $\Omega$, the period of the medium, so that
$\delta = \Omega/\lambda$ is a small parameter.
Here we apply the same homogenization process to the nonlinear system
\eqref{2D psystem conservation form} to derive homogenized equations 
with $\order(\delta^2)$ corrections.
Details of the homogenization process are deferred to Appendix \ref{appendix: homogenized equations}.
The resulting homogenized system is
\begin{subequations} \label{homogenized_system}
\begin{align}
 K_h^{-1}\sigma_t-\left(\sigma+1\right)\left(u_x+v_y\right) & = \delta^2\alpha_1\left[\left(\sigma+1\right)\left(u_{xyy}+v_{yyy}\right)+2\sigma_y\left(u_{xy}+v_{yy}\right)\right] \nonumber \\
 											& \quad +\delta^2\alpha_2\left[\left(\sigma+1\right)\left(u_{xxx}+v_{xxy}\right)+2\sigma_x\left(u_{xx}+v_{xy}\right)\right] \nonumber \\
 											& \quad +\delta^2\alpha_3\sigma_y\left(u_{xy}+v_{yy}\right),\\
 \rho_h u_t-\sigma_x & = \delta^2 \beta_1\sigma_{xyy}+\delta^2 \beta_2\sigma_{xxx},\\
 \rho_m v_t-\sigma_y & = \delta^2 \gamma_1\sigma_{yyy}+\delta^2 \gamma_2 \sigma_{xxy},
\end{align}
\end{subequations}
where the values of the coefficients $\alpha, \beta, \gamma$ depend on the the material
coefficient functions $K(y), \rho(y)$.
The subscripts $m$ and $h$ denote the arithmetic and harmonic average, respectively:
\begin{align*}
\rho_m & = \int_0^1 \rho(y) dy, & \rho_h & = \left(\int_0^1 \frac{1}{\rho(y)} dy\right)^{-1}.
\end{align*}

Observe that the $\order(1)$ terms in \eqref{homogenized_system} represent
a straightforward averaging of the variable-coefficient equations.  Meanwhile,
the $\order(\delta^2)$ terms introduce dispersion.
For the layered medium \eqref{discontinuous medium}, 
it is possible to obtain closed-form expressions for the dispersive term
coefficients:
\begin{subequations} \label{layered_medium_coefficients}
\begin{align}
\gamma_1 & = \lambda^2 \left(Z_A^2-Z_B^2\right)\frac{\left(\rho_A-\rho_B\right)}{192K_m\rho_m^2}
& \gamma_2 & = \lambda^2 \left(c_A^2-c_B^2\right)\frac{\left(\rho_A-\rho_B\right)}{192K_m}, \\
\beta_1 & = \lambda^2 \left(Z_A^2-Z_B^2\right)\frac{\left(\rho_B-\rho_A\right)}{192K_m\rho_m^2}
& \beta_2 & = \lambda^2 \left(c_A^2-c_B^2\right)\frac{\left(\rho_B-\rho_A\right)}{192K_m},\\
 \alpha_1 & = \lambda^2 \left(Z_A^2-Z_B^2\right)\frac{\left(K_A-K_B\right)}{192K_m^2\rho_m} &
 \alpha_2 & = \lambda^2 \left(c_A^2-c_B^2\right)\frac{\left(K_A-K_B\right)\rho_m}{192K_m^2},\\
 \alpha_3 & = \lambda^2 \frac{\left(\rho_A-\rho_B\right)^2}{192\rho_m^2}.
\end{align}
\end{subequations}
Notice that $\alpha_1, \beta_1, \gamma_1$ vanish when $Z_A=Z_B$, whereas
$\alpha_2, \beta_2, \gamma_2$ vanish when $c_A = c_B$.  These properties
will play an important role in what follows.

\subsection{Normally-propagating plane waves}
For initial data that do not vary in $x$, solutions to \eqref{homogenized_system}
are plane waves traveling in the normal direction (parallel to the $y$-axis).
For such waves, system \eqref{homogenized_system} simplifies to:
\begin{subequations} \label{homog_normal}
\begin{align}
 K_h^{-1}\sigma_t-\left(\sigma+1\right)v_y & = \delta^2 \alpha_1\left(\sigma+1\right)v_{yyy}+\delta^2\left(\alpha_3+2\alpha_1\right)\sigma_y v_{yy},\\
 \rho_m v_t-\sigma_y & = \delta^2 \gamma_1 \sigma_{yyy}.
\end{align}
\end{subequations}
The corresponding one-dimensional problem is studied extensively in \cite{leveque2003},
and \eqref{homog_normal} is equivalent (up to $\order(\delta^2)$) to the
approximation derived there.
This homogenized system (and the original variable-coefficient equation) possesses
solitary wave solutions, due to the combination of nonlinearity and an effective 
dispersion that arises due to reflection.  We do not pursue this case further here,
except to observe that most of the dispersive terms vanish when the linearized impedance
is constant (i.e., when there is no reflection).  

\subsection{Transversely-propagating plane waves\label{sec:homog_transverse}}
For initial data that do not vary in $y$, solutions to \eqref{homogenized_system}
are plane waves traveling in the transverse direction (parallel to the $x$-axis).
For such waves, system \eqref{homogenized_system} simplifies to:
\begin{subequations} \label{homog_transverse}
\begin{align}
 K_h^{-1} \sigma_t-\left(\sigma+1\right)u_x & = \delta^2 \alpha_2 \left[\left(\sigma+1\right)u_{xxx}+2\sigma_x u_{xx}\right],\\
 \rho_h u_t-\sigma_x & = \delta^2 \beta_2 \sigma_{xxx}.
\end{align}
\end{subequations}
As our introductory experiments (see Figure \ref{fig:intro_experiments})
suggest, this system possesses solitary wave
solutions as long as the sound speed is not constant.  On the other hand,
if the linearized sound speed is constant then all the dispersive term
coefficients vanish.  This is because the effective dispersive
mechanism in this case is that of diffraction, which occurs only if the sound speeds
differ \cite{QK2013}.  In the absence of diffraction, nonlinearity
leads to shock formation, as observed in Figure \ref{fig:nonlinear_constant}.

\begin{figure}
\begin{centering}
  \includegraphics[scale=0.35]{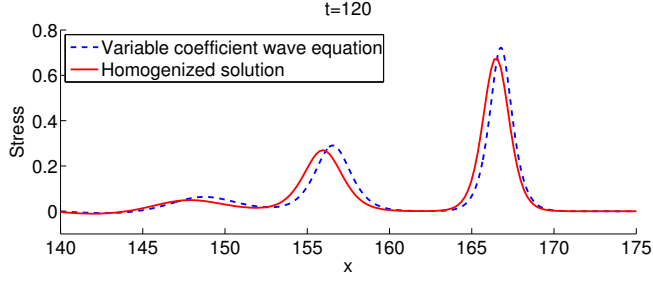}
\par\end{centering}
\caption{Solution of one-dimensional homogenized equations
\eqref{homog_transverse} (solid red) versus $y$-averaged solution of the
two-dimensional variable-coefficient wave equation \eqref{p-system} (dashed
blue).\label{fig:homog_transverse}}
\end{figure}

In Figure \ref{fig:homog_transverse}, we compare the numerical solution of 
\eqref{homog_transverse} with that of the variable coefficient 2D wave equation
\eqref{p-system}, arithmetically averaged in $y$.
The initial condition is
\begin{align} \label{gaussian}
\sigma_0(x,y) & = e^{-x^2/10}, & u_0 = v_0 & = 0.
\end{align}
The solitary wave solutions of \eqref{homog_transverse}
are a reasonably good approximation to the solutions of \eqref{p-system}, and
could be improved by including higher-order terms.

\subsection{Stationary solutions of the homogenized equations\label{sec: homogenized diffractons}}
In this section, we assume a traveling wave 
solution for \eqref{homog_transverse}
and derive an ODE for the shape of a homogenized $\order(\delta^2)$
diffracton. Afterwards, following Section 10 of \cite{ketchesonphdthesis}, 
we use this ODE to find a lower limit for the speed of a diffracton.

Consider system \eqref{homog_transverse}
and assume a traveling wave solution of the form 
$\sigma+1=W(x-Vt)$ and $u=U(x-Vt)$, where $V$ is the speed 
of the traveling wave. After combining the resulting equations
and dropping terms of $\order(\delta^4)$ we get: 
\begin{align} \label{ODE for traveling wave homogenized diffractons}
 W' & = \frac{1}{M^2} \left[WW' + \delta^2(\alpha_2+\beta_2)WW'''  + 2\alpha_2\delta^2W'W''\right].
\end{align}
Here we have introduced the mach number $M=V/\ceff$,
where $\ceff=\sqrt{K_h/\rho_m}$ is the speed at which small-amplitude,
long-wavelength perturbations travel in the transverse direction.
For the layered medium \eqref{discontinuous medium}, the 
coefficients $\alpha_2$ and $\beta_2$ are given by 
\eqref{layered_medium_coefficients}
and their sum is non-negative:
\begin{align}
  \alpha_2+\beta_2 = \lambda^2 \frac{(c_A^2-c_B^2)^2}{192 K_m^2/(\rho_m\rho_h)} \ge 0.
\end{align}

%

Numerically integrating \eqref{ODE for traveling wave homogenized diffractons}
with an appropriate velocity $V$ yields solitary waves nearly identical to the
homogenized diffractons of Figure \ref{fig:homog_transverse}.

Larger-amplitude waves move faster than $\ceff$, while short-wavelength, small-amplitude waves
move more slowly (due to diffractive dispersion).  Since diffractons are nonlinear
waves with wavelength on the order of a few material layers, it is not clear
{\em a priori} whether their speed should be larger or smaller than $\ceff$.
Here we show that the homogenized equations indicate that diffractons move
faster than $\ceff$; this is confirmed experimentally in Section \ref{sec:speed-amplitude}.

Integrate by parts  
\eqref{ODE for traveling wave homogenized diffractons} 
and let $w_1=W$ and $w_2=W'$ to get:
\begin{subequations}
\begin{align}
 w_1' &= w_2, \\
 w_2' &= -\frac{(\alpha_2-\beta_2)w_2^2}{(\alpha_2+\beta_2)w_1} - 
 		\frac{w_1}{2\delta^2(\alpha_2+\beta_2)} + 
		\frac{M^2}{\delta^2(\alpha_2+\beta_2)}\left(1-\frac{2-M^{-2}}{2w_1} \right),
\end{align}
\end{subequations}
The equilibrium points are $(1, 0)$ and $(2M^2-1, 0)$ and the Jacobian is:
\begin{align}
 J & = \left[
 	  \begin{array}{cc}
		0 & 1\\
		-\frac{1-2M^2+w_1^2-2(\alpha_2-\beta_2)\delta^2w_2^2}{2(\alpha_2+\beta_2)\delta^2w_1^2} 
		& \frac{-2(\alpha_2-\beta_2)w_2}{(\alpha_2+\beta_2)w_1} 
	  \end{array}
	 \right],
\end{align}
whose eigenvalues at $(1, 0)$ and $(2M^2-1, 0)$ are: 
\begin{subequations}
 \begin{align}
  	&  \lambda_l =\pm \frac{\sqrt{M^2-1}}{\delta\sqrt{\alpha_2+\beta_2}}, 
  	&  \lambda_r =\pm \frac{\sqrt{1-M^2}}{\delta\sqrt{(\alpha_2+\beta_2)(2M^2-1)}},
\end{align}
\end{subequations}
respectively. 
The boundary conditions for a solitary wave are 
$W \rightarrow 1$ and $W', W'', W'''\rightarrow 0$ as $|x-Vt| \rightarrow \infty$. 
The boundary condition at $|x-Vt|\rightarrow \infty$ corresponds to 
the equilibrium point $(1, 0)$; thus, diffractons correspond 
to homoclinic connections for this point. A homoclinic connection 
occurs if the equilibrium point $(1, 0)$ is a saddle and $(2M^2-1, 0)$ 
is a center. This happens only when $|M|>1$, so the homogenized 
diffractons are ``supersonic''. This property also holds for stegotons~\cite{ketchesonphdthesis}.

\section{Properties and dynamics of diffractons\label{sec:properties}}
In this section we investigate the properties of diffractons: their stability, shape, 
scaling properties, speed-amplitude relation, and interactions.

\subsection{Long-time stability}
We have already seen that numerical solutions of both \eqref{p-system}
and \eqref{homogenized_system} may lead to the appearance of diffractons
from general initial data.  Indeed, it seems that diffractons are globally
attracting solutions.  To further investigate their long-time behavior we take
a single diffracton from the solution of \eqref{p-system} as initial data
and propagate it to $t=600$; the diffracton travels more than 600 units in
space.  Let $x_m(t)$ denote the grid location of the diffracton peak at time $t$.
We compute the maximum relative difference between the solution at $t=0$ and the (re-centered)
solution at time $t$:
\begin{equation} 
D = \max_t \left( \frac{\left\Vert \sigma(x-x_m(0),y,t=0)-\sigma(x-x_m(t),y,t)\right\Vert _{2(x,y)}}{\left\Vert \sigma(x-x_m(0),y,t=0)\right\Vert _{2(x,y)}}\right).
\end{equation}
We consider two different grids: on a grid with $\Dx = \Dy = 1/16$ the maximum
difference is $D=4.6\%$; with $\Dx = \Dy = 1/32$, it is $D=2.2\%$.
Because we have taken $x_m(t)$ as simply the nearest grid point to the maximum,
first order convergence is expected.  Hence these results suggest that the
computed solution has a constant shape, up to numerical error.





\subsection{Speed-amplitude relation} \label{sec:speed-amplitude}
\begin{figure}
\begin{centering}
  \includegraphics[scale=0.35]{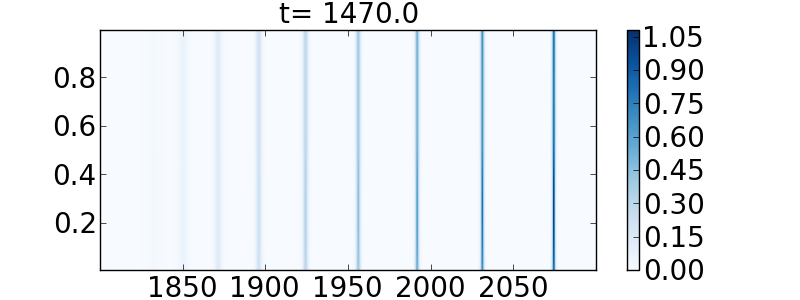}
  \includegraphics[scale=0.35]{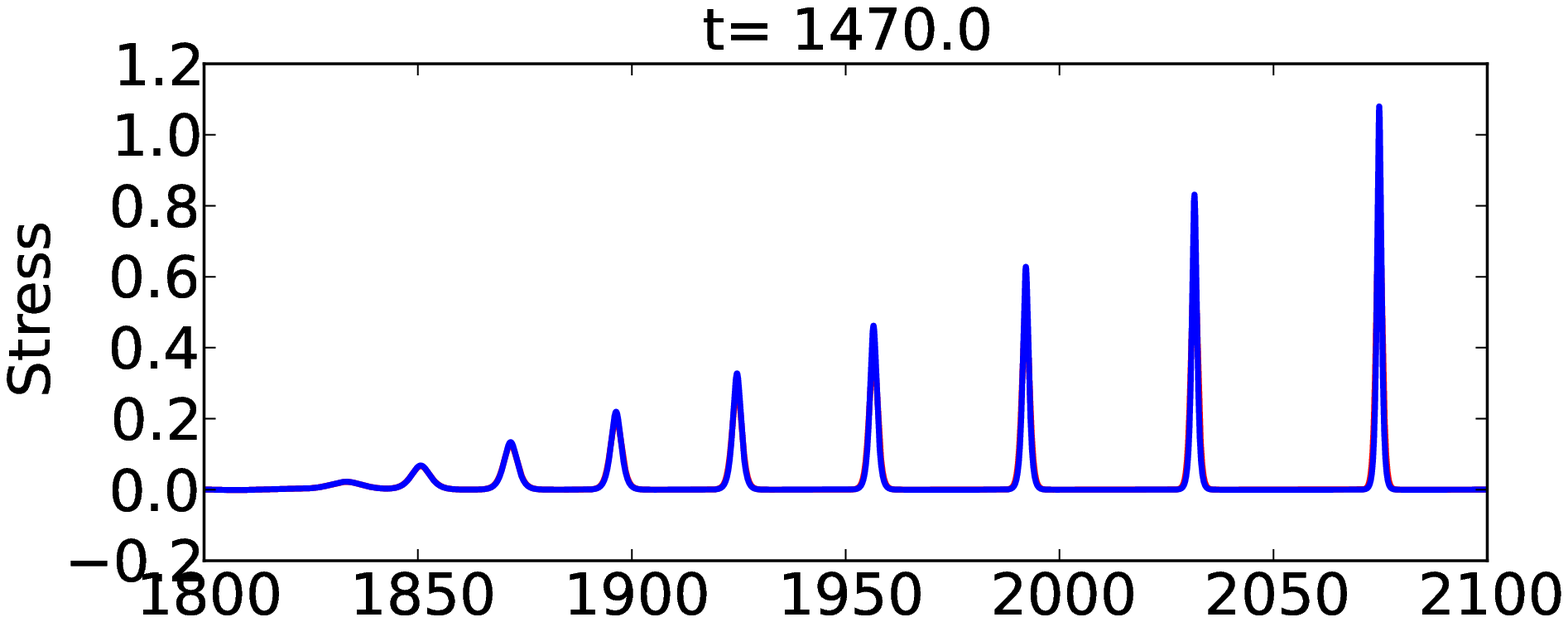}
\par\end{centering}
\caption{Formation of diffractons used to study scaling properties and 
the speed-amplitude relation. Left: stress as a function of $x$ and $y$.
Right: slices at the middle of material A (blue) and B (red). 
\label{fig: many diffractons}}
\end{figure}

There is a simple relationship between the $x$-momentum amplitude
$A = \max_{x,y} \rho|u|$ of a diffracton and its speed, $V$.
In order to demonstrate this, we take a very broad initial condition:
\begin{align}
\sigma_0(x,y) & = e^{-x^2/100}, & u_0 = v_0 & = 0.
\end{align}
The solution, which evolves into eight separate diffractons, is shown in
Figure \ref{fig: many diffractons}.

\begin{figure}
\begin{centering}
  \includegraphics[scale=0.3]{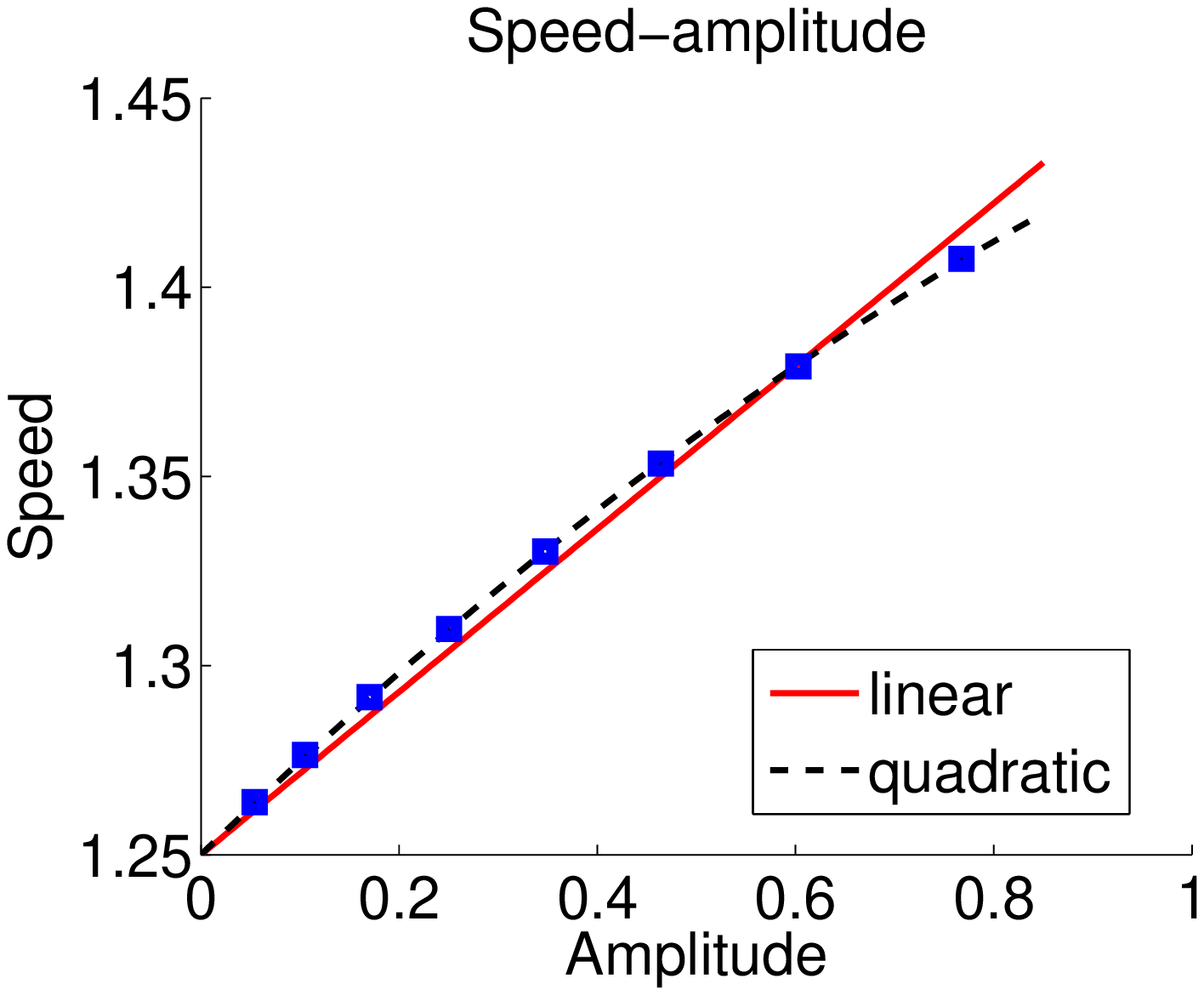}
  \includegraphics[scale=0.3]{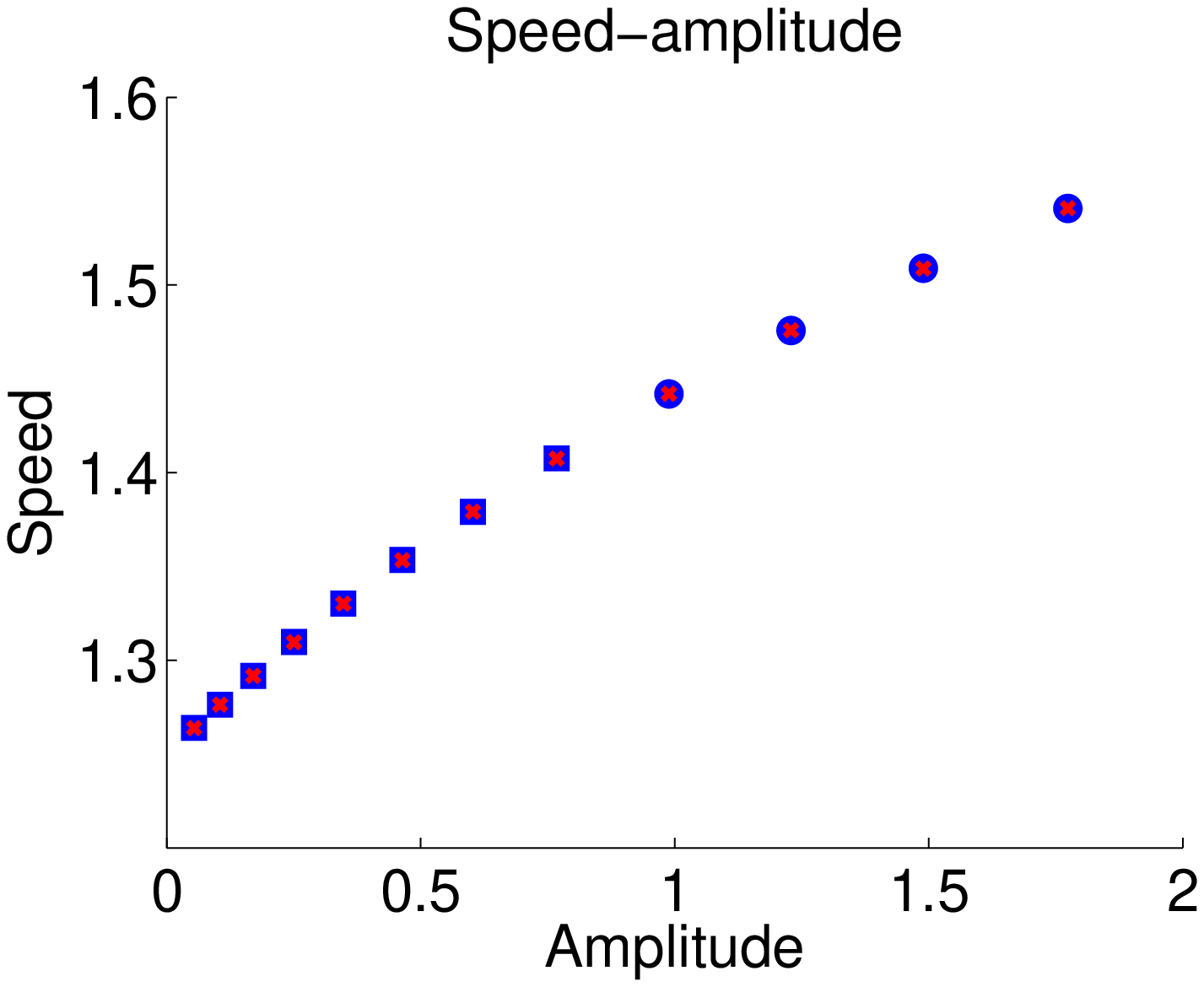}
\par\end{centering}
\caption{Speed-amplitude relationship for diffractons.
The blue circles denote measurements.
{\bf Left:} Linear (red) and quadratic (black) least-squares fits.
{\bf Right:} Comparison with values predicted by
\eqref{speed of solitary waves} (red crosses). 
\label{fig: speed-amplitude}}
\end{figure}

We isolate each diffracton and propagate it up to $t=100$ to measure its speed.
The blue squares in Figure \ref{fig: speed-amplitude} (left)
show the measured speeds versus the amplitude $A$. In addition, 
we show linear (red) and quadratic (black) least-squares fitted curves
(constrained to pass through the known value $V=\ceff=0.25$ for
zero-amplitude waves). 
It is clear that the speed-amplitude relation is nonlinear.
In this respect, diffractons are different from stegotons, which
appear to have a linear speed-amplitude relation~\cite{leveque2003}.
Many other classes of solitary waves are known to have a nonlinear 
speed-amplitude relationship; see for instance 
\cite{vlasenko2000structure, akers2008model, duran2013galilean}. 

We have found (empirically) the very simple relation
$V \approx \frac{\max{\sigma(x,y)}}{A}$.  Indeed, we even have
the relation
\begin{equation} \label{speed of solitary waves}
V \approx \frac{{\sigma}}{\rho|u|}.  
\end{equation}
Here the numerator and denominator are functions of $x$ and $y$,
but the ratio is essentially constant whenever $|u|$ is much larger
than roundoff. 
In Figure \ref{fig: speed-amplitude} (right) we show the speed predicted by 
\eqref{speed of solitary waves} (red crosses) and the measured
speed (blue squares) for the first eight diffractons in figure 
\ref{fig: many diffractons} and for four larger 
diffractons constructed obtained by a scaling procedure described in
Section \ref{sec: scaling} (blue circles).  The values are indistinguishable.

In experiments with media obtained by other choices of $(K_A, K_B, \rho_A, \rho_B)$
we have found that the speed of each solitary wave is always given to high
accuracy by the ratio $\max_x \sigma / \max_x (\rho|u|)$, and this value is
independent of $y$.  But for other media it is not always true that this value
is the same for all $x$ within a given solitary wave.

\subsection{Scaling} \label{sec: scaling}
Many one-dimensional solitary waves (such as the soliton solutions of the KdV equation)
are known to have a shape identical or similar to that of
the function $\mathop{sech}^2(x)$; furthermore, the width of a solitary wave
often varies inversely with the square root of its amplitude.  These properties were found
to (approximately) hold for stegotons in \cite{leveque2003}.
The one-dimensional cross-sections of a diffracton (with $y$ fixed) approximately
satisfy these properties,
although the amplitude of the cross-section is different for each $y$ value.

\begin{figure}
\begin{centering}
\includegraphics[scale=0.30]{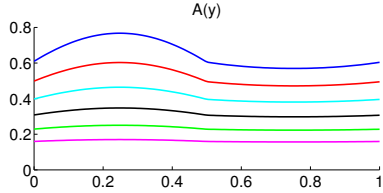}
\par\end{centering}
\caption{Amplitude function $A(y)$ for each of the first six diffractons in Figure \ref{fig: many diffractons}.\label{fig:amplitude}}
\end{figure}

\begin{figure}
\centerline{
\subfigure[x-momentum\label{fig:scaling_u}]{\includegraphics[scale=0.3]{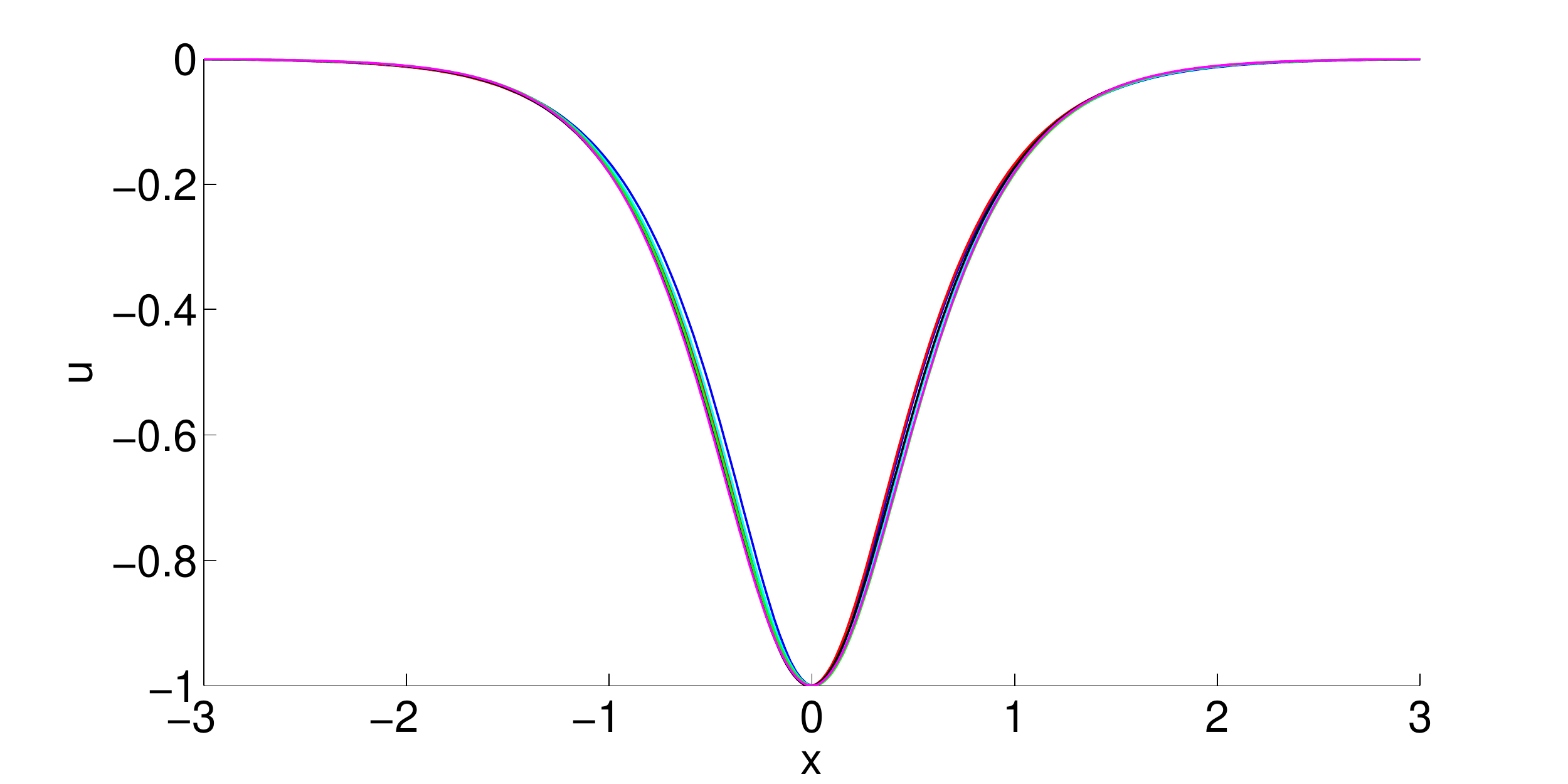}}
\subfigure[Stress\label{fig:scaling_stress}]{\includegraphics[scale=0.3]{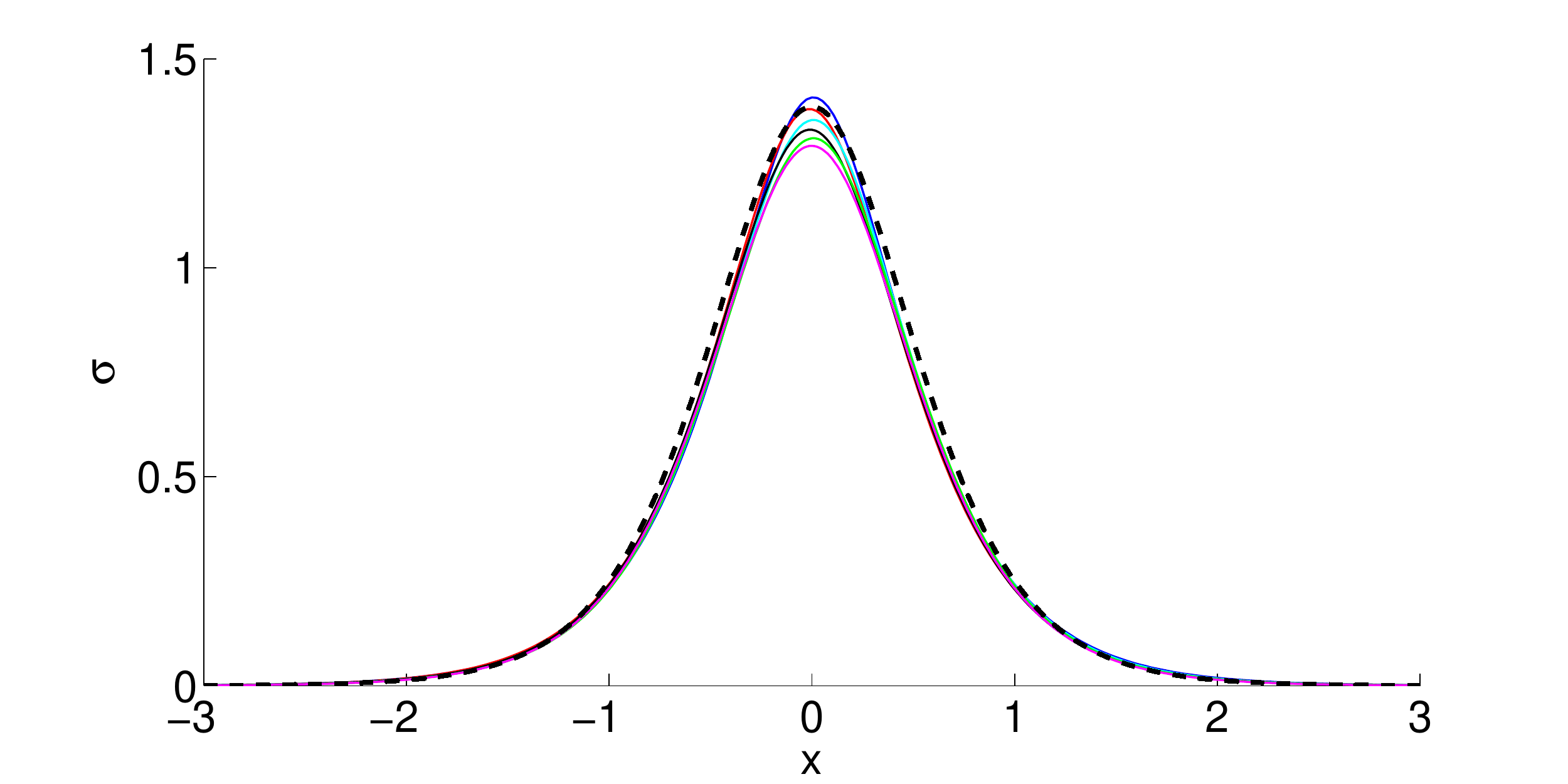}}}
\centerline{
\subfigure[y-momentum\label{fig:scaling_v}]{\includegraphics[scale=0.3]{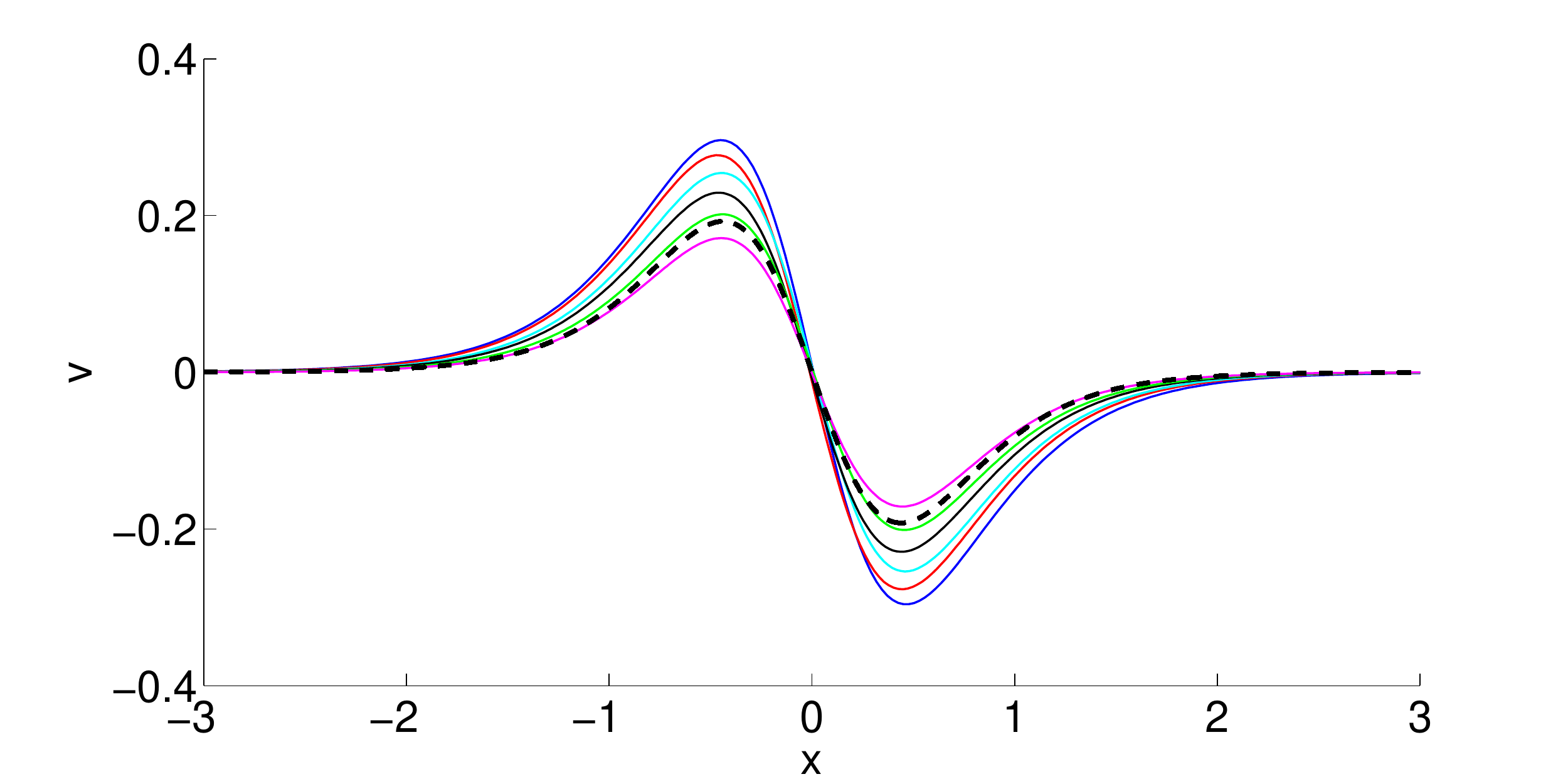}}
\subfigure[y-momentum, alternative scaling\label{fig:scaling_v_alt}]{\includegraphics[scale=0.3]{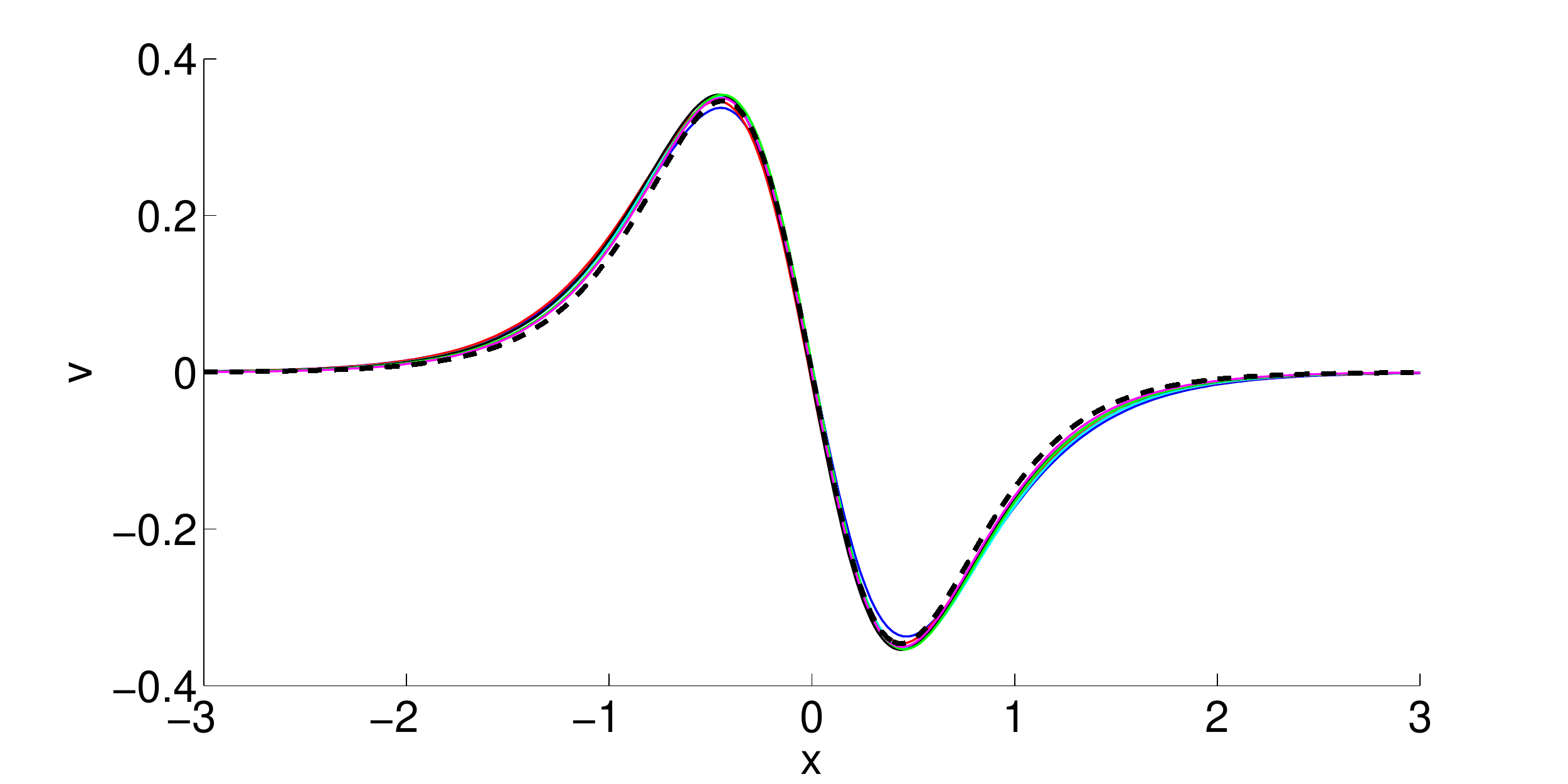}}}
\caption{Slices at $y=1/8$ of rescaled velocities and stress.  The dashed
lines represent a $\mathop{sech}^2$ curve (figure (b)) and its derivative (figures (c-d)).\label{fig:scaling}}
\end{figure}

For this section we define the amplitude as a function of $y$, and we also 
consider the location of the peak amplitude:
\begin{align*}
A(y) & = \max_x \rho(x,y)|u(x,y)|, \\
x_m(y) & = \textup{argmax}_x \rho(x,y) |u(x,y)|.
\end{align*}
We observe that $x_m$ is in fact independent of $y$.
The amplitude function for various diffractons is plotted in Figure \ref{fig:amplitude}.
Then the stress and $x$-momentum of each cross-section of different diffractons
appear to have the same shape under the transformation:
\begin{align*}
f(x,y) & \to \frac{1}{A(y)} f(\sqrt{A(y)}(x-x_m))
\end{align*}
Of course, the transformed stress peak amplitudes of different waves will not
be equal, since they are just the velocities of the corresponding diffractons.
In Figures \ref{fig:scaling_u} and \ref{fig:scaling_stress}, we have plotted
these transformed values for the leading six diffractons from Figure 
\ref{fig: many diffractons}, along the line
$y=1/8$.  Slices at other $y$-values look similar.  
In Figure \ref{fig:scaling_stress} we have also plotted (dashed line) a $\mathop{sech}^2$
function with amplitude and width fitted to the data.

Figure \ref{fig:scaling_v} shows the values of $\rho v$ under the same
transformation.  The locations of the extrema of the different curves 
coincide, but the amplitudes do not.  We have also plotted (dashed line)
the function $\partial_x \mathop{sech}^2$, again with amplitude and width fitted
to the data.  Finally, in Figure \ref{fig:scaling_v_alt}, we plot the $y$-momentum
under the empirically-determined transformation
\begin{align*}
f(x,y) & \to \frac{1}{A(y)^{1.4}} f(\sqrt{A(y)}(x-x_m)),
\end{align*}
which seems to scale the amplitudes almost equally.


\subsection{Interaction of diffractons}
We now investigate the behavior of colliding diffractons.
The diffractons used for these experiments are shown in the 
top-left panel of Figure \ref{fig: Co-propagating-collision}.
All plots shown are $y$-slices of the stress at the middle 
of material A (blue) and material B (red). 

\subsection{Co-propagating collision}
\begin{figure}
\begin{centering}
  \includegraphics[scale=0.35]{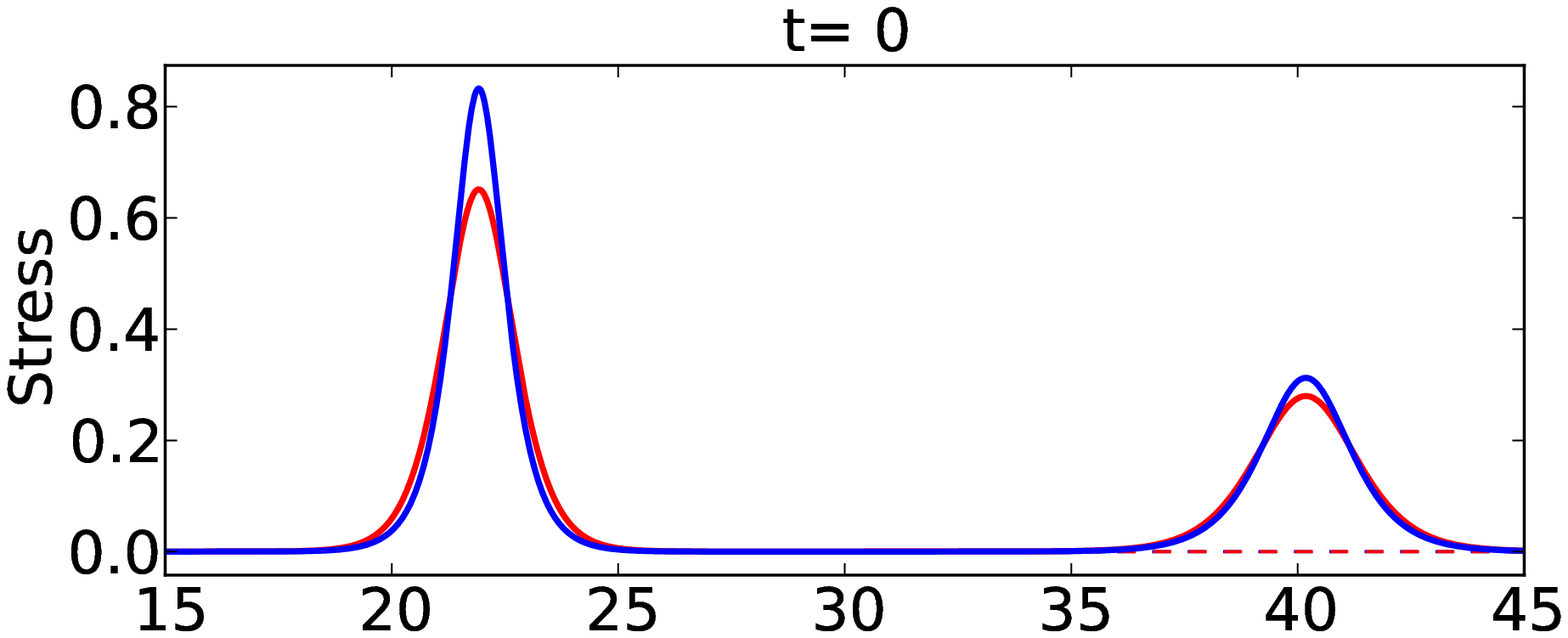}
  \includegraphics[scale=0.35]{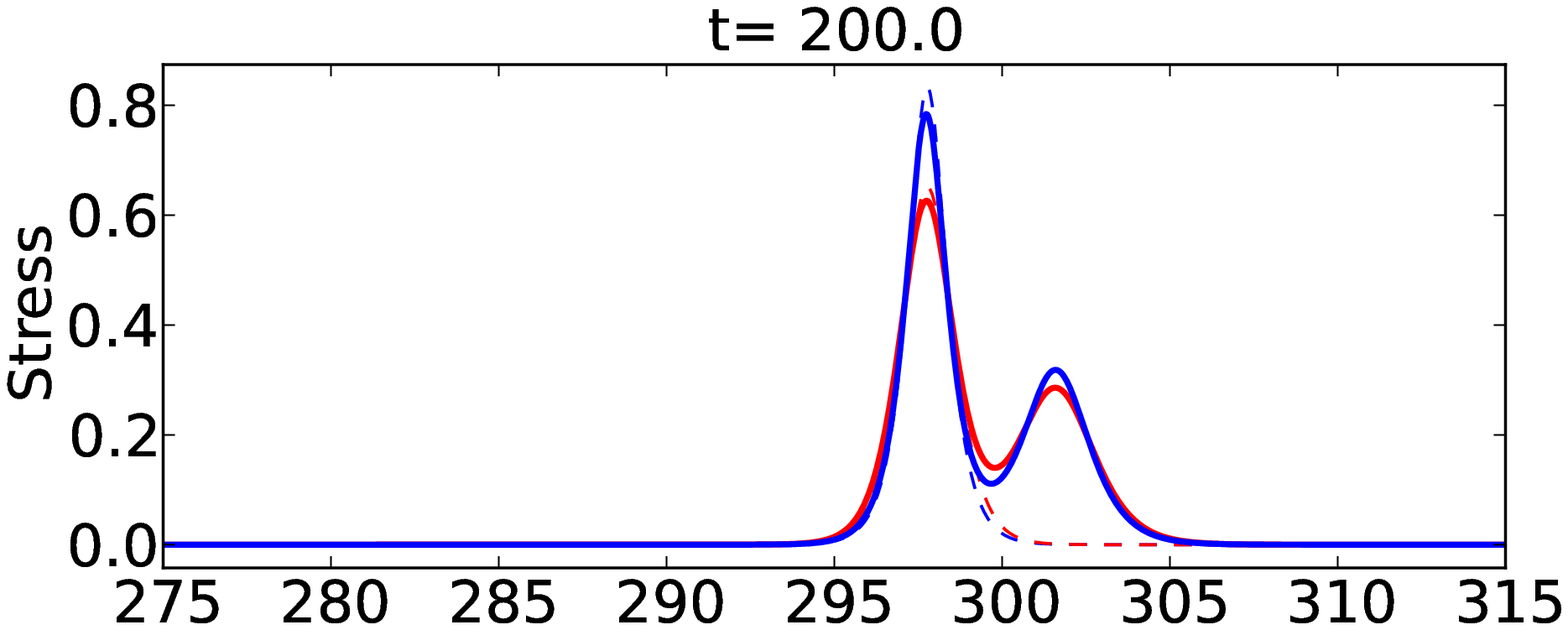}

  \includegraphics[scale=0.35]{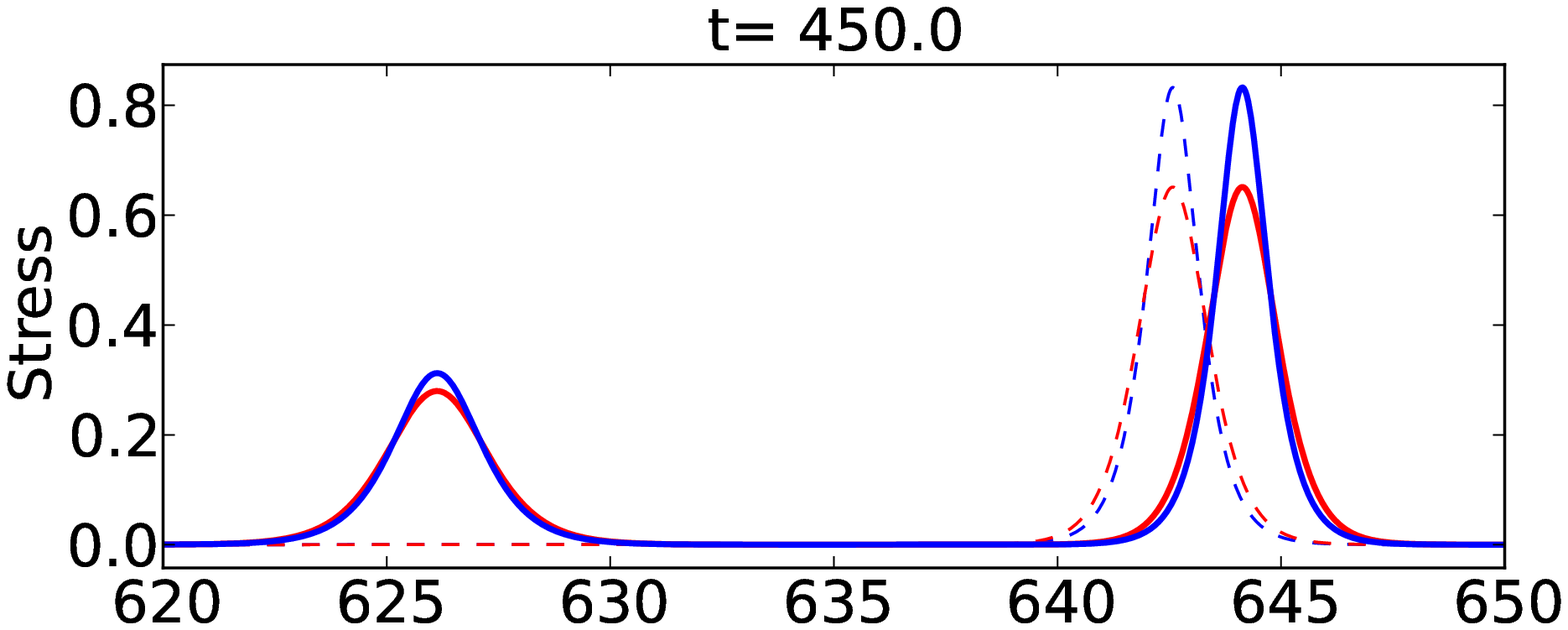}
  \includegraphics[scale=0.35]{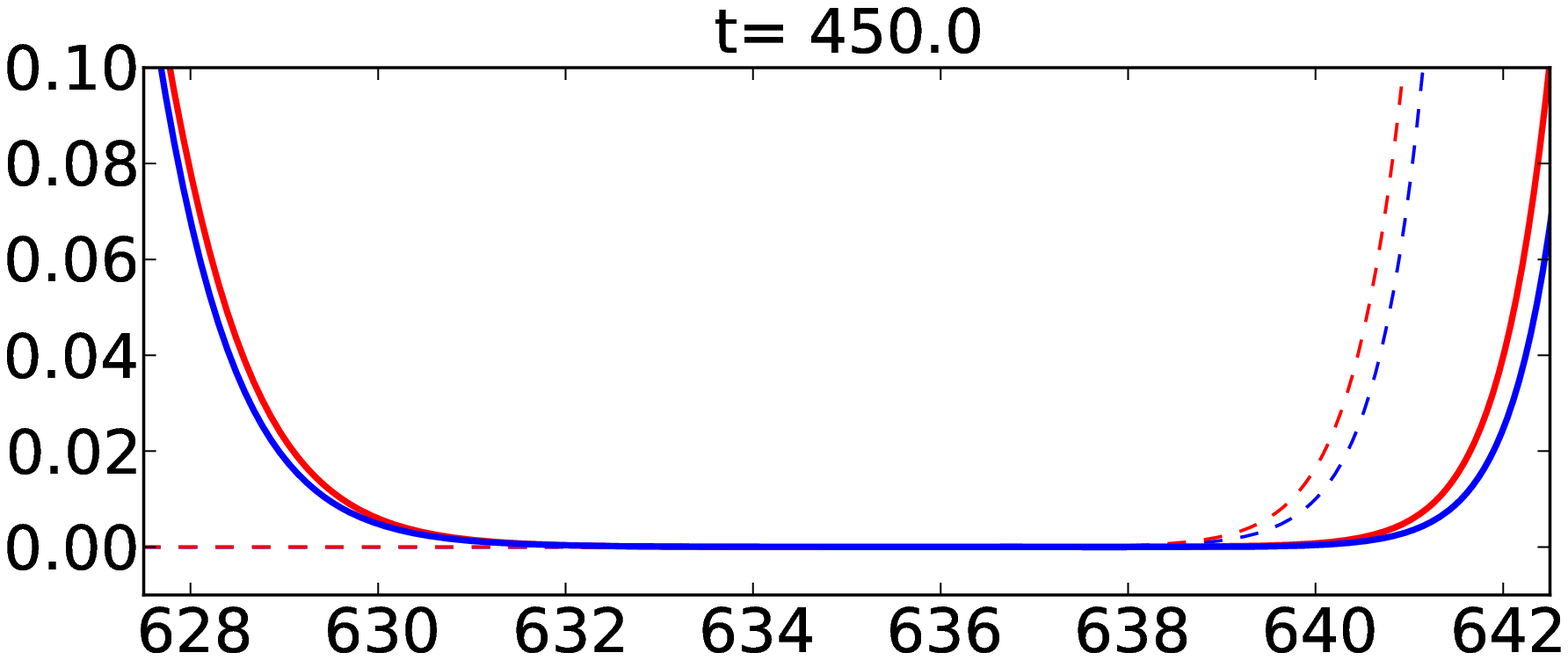}
\par\end{centering}
\caption{Co-propagating collision at different times and a close-up 
to the tail of the diffractons after the interaction (bottom right).\label{fig: Co-propagating-collision}}
\end{figure}

In this scenario, both waves are moving to the right.
Figure \ref{fig: Co-propagating-collision} shows the stress at
different times during the interaction (solid line).  For comparison,
the dashed line shows the propagation of the taller wave
by itself.  As in most solitary wave interactions, a clear phase shift is exhibited.
No oscillations are visible  after the interaction, as
shown in the lower-right panel.
This suggests that the interaction is elastic, which is 
often the case in co-propagating collisions of solitons 
and other solitary waves \cite{hirota1971exact, zabusky1965interaction,wu1998nonlinear}.

\subsection{Counter-propagating collision }
\begin{figure}
\begin{centering}
  \includegraphics[scale=0.35]{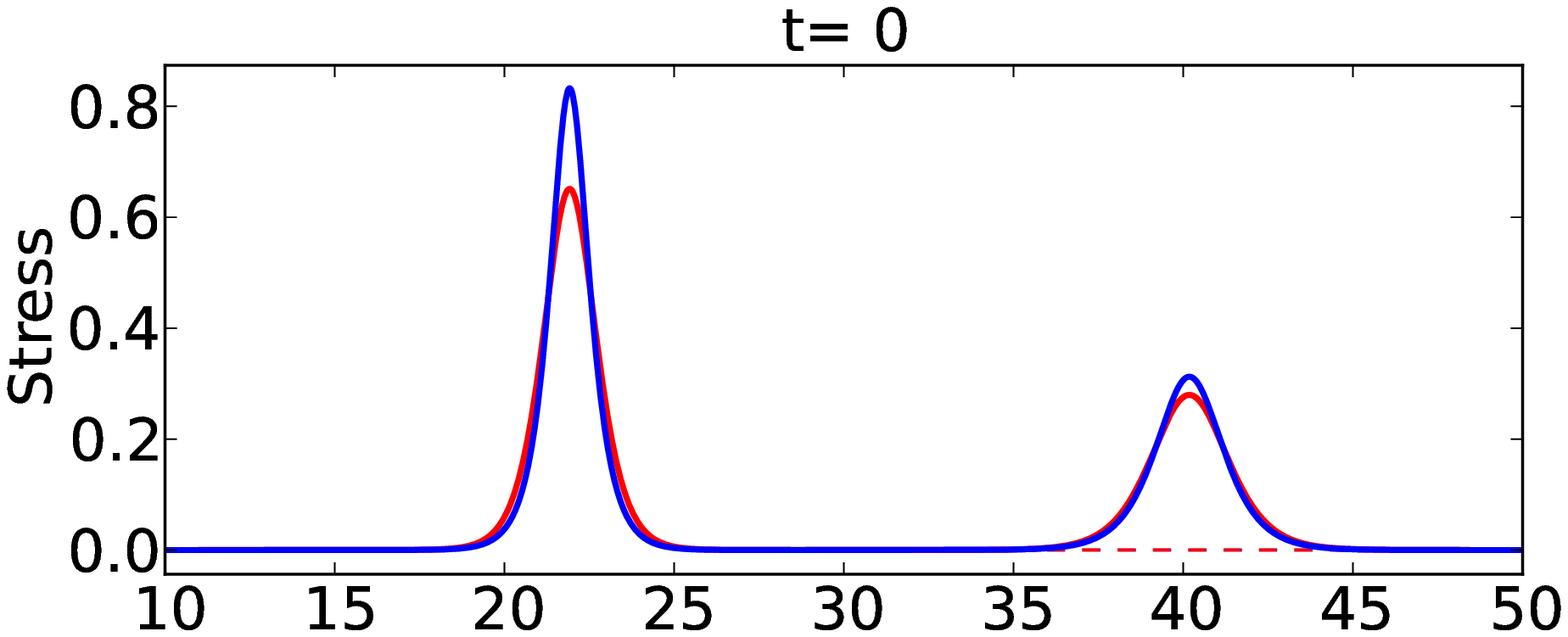}
  \includegraphics[scale=0.35]{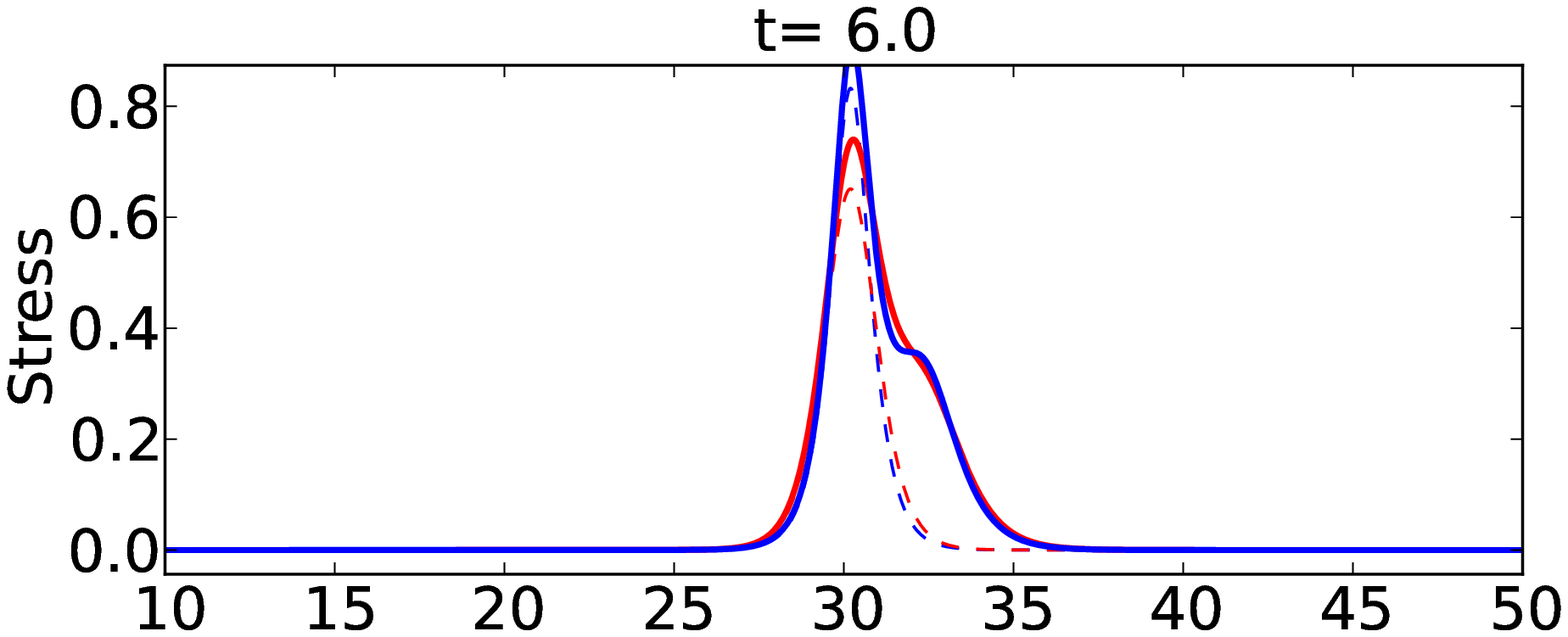}

  \includegraphics[scale=0.35]{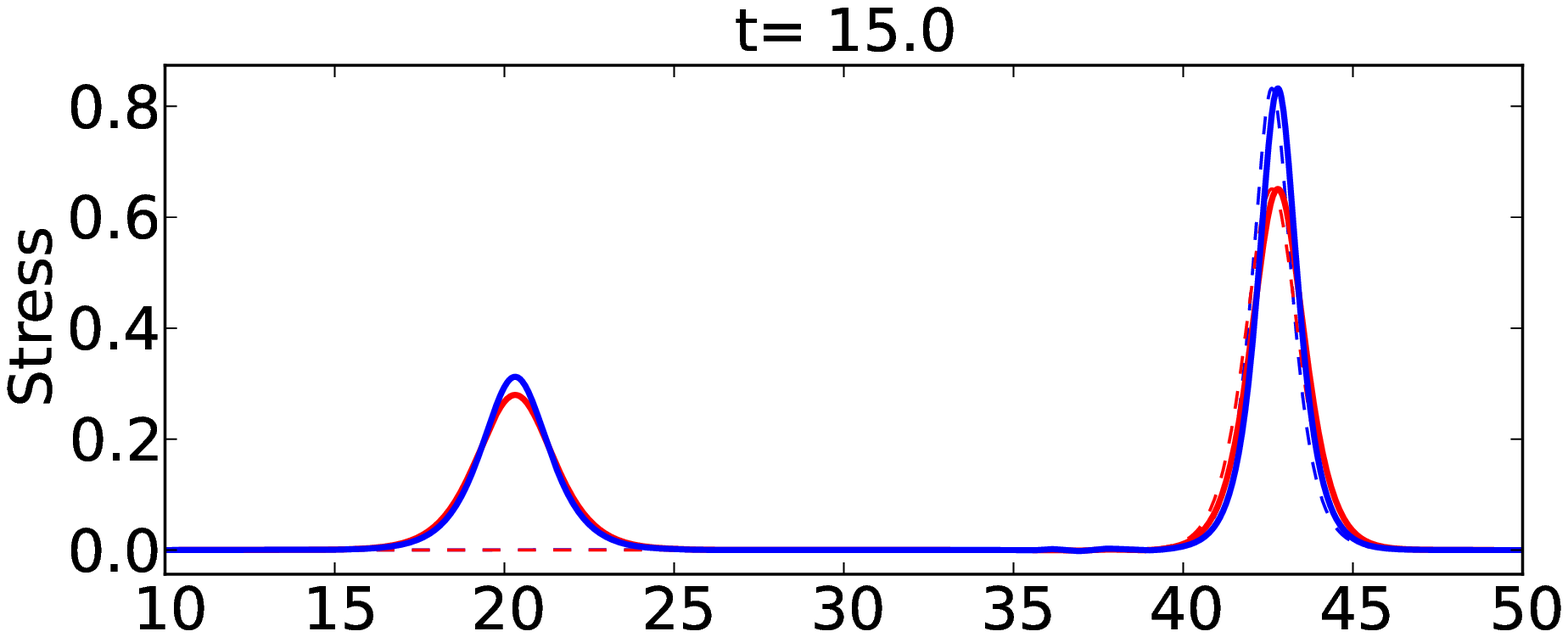}
  \includegraphics[scale=0.35]{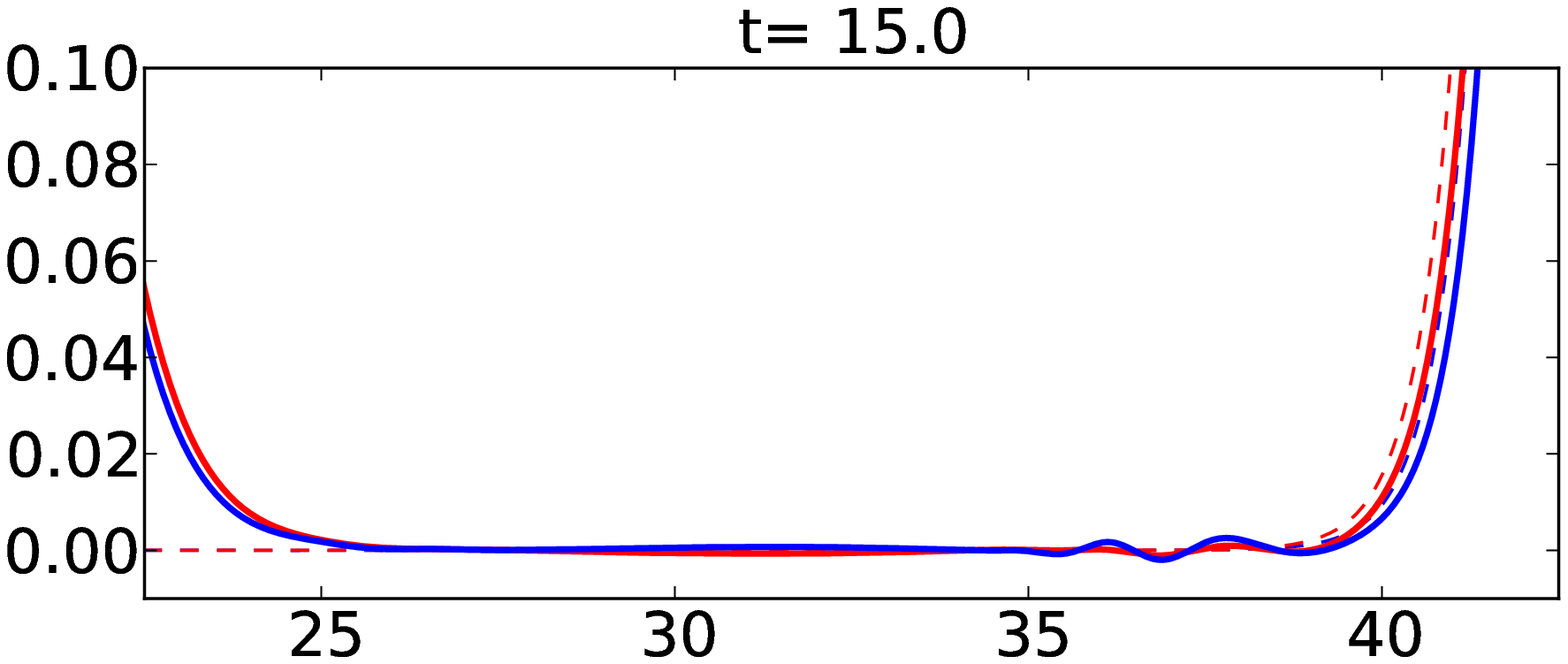}
\par\end{centering}
\caption{Counter-propagating collision at different times and a close-up of the tail 
of the diffractons after the interaction (bottom right).\label{fig: Counter-propagating-collision}}
\end{figure}

In Figure \ref{fig: Counter-propagating-collision} we consider the same
solitary waves in the same initial locations, but we
negate the velocity fields $u,v$ of the shorter wave to make it propagate to
the left while the taller wave propagates to the right.
This time there is barely a trace of phase shift; this is typical
when the interaction time is so short.  Oscillations {\em are}
seen after the collision.  

\begin{figure}
\begin{centering}
  \includegraphics[scale=0.5]{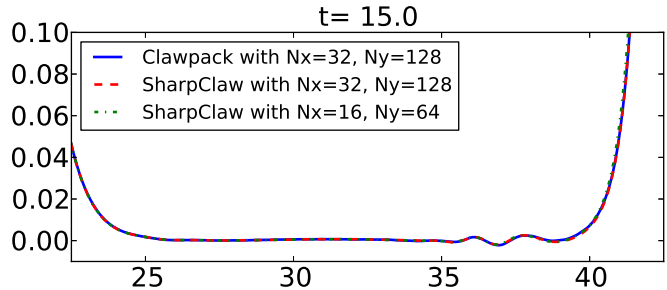}
\par\end{centering}
\caption{Close-up of the tail of the diffractons after a counter-propagating collision
using SharpClaw on a grid with $N_x=32$ and $N_y=128$ grid points per unit
in $x$ and $y$ respectively (solid blue line), Clawpack on the same
grid (dashed red line) and SharpClaw on a coarser grid with $N_{x}=16$
and $N_{y}=64$ (dotted greed line). In this case, just the $y$-slice
at the middle of material A is shown. \label{fig: counter-propagating comparison}}
\end{figure}

To investigate whether the oscillations
are numerical or physical, in Figure \ref{fig: counter-propagating comparison}
we repeat the same simulation on a grid with half as many points using SharpClaw and on
the same grid using Clawpack with a TVD slope-limiter.
Essentially no change in the solution is observed, strongly suggesting
that the oscillations are physical (i.e., the counter-propagating
collision is not elastic). This behavior has been observed for other
solitary waves; e.g., in \cite{su1980head,mirie1982collisions,byatt1989head}. 

\section{Generalizations\label{sec:generalizations}}
The formation of solitary waves in general results from a balance between
dispersion and nonlinearity.  This section demonstrates that diffracton
solutions exist under a broad range of scenarios.

\subsection{Smoothly-varying medium}
Effective dispersion due to diffraction occurs not only in the piecewise-constant
media we have focused on, but more generally in any periodic medium with
variable sound speed.
As an example, we consider a sinusoidally-varying medium, with coefficients
\begin{subequations} \label{sinusoidal medium}
\begin{align}
  K(y) & = \frac{K_A+K_B}{2} +
                \frac{K_A-K_B}{2} \sin\left(2\pi y\right), \\
  \rho(y) & = \frac{1}{K(y)}.
\end{align}
\end{subequations}
We solve the homogenized equations \eqref{homog_transverse} for a transverse
perturbation.  
We take the material parameters \eqref{zmatched} and as initial data the
Gaussian stress perturbation \eqref{gaussian}.
The coefficients in this case are different from those for 
the piecewise medium; see the Appendix.
Figure \ref{fig: sin diffracton} shows the solution at $t=120$; solitary wave
solutions are again observed.  
\begin{figure}
\begin{centering}
  \includegraphics[scale=0.35]{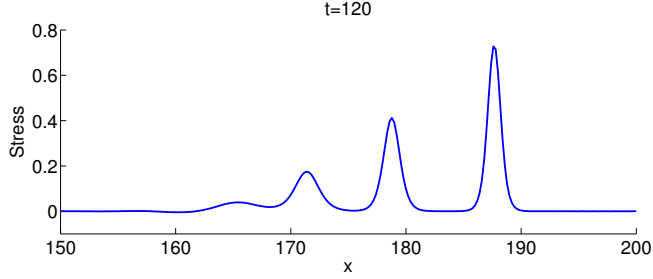}
\par\end{centering}
\caption{Homogenized diffractons for a sinusoidal medium given by 
\eqref{sinusoidal medium}. \label{fig: sin diffracton}}
\end{figure}

\subsection{Quadratic nonlinearity}
\begin{figure}
\begin{centering}
  \includegraphics[scale=0.35]{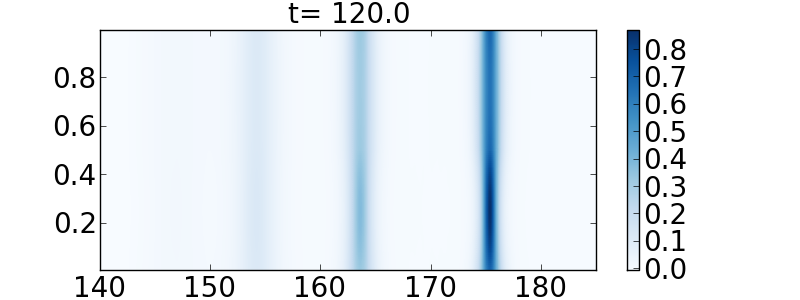}
  \includegraphics[scale=0.35]{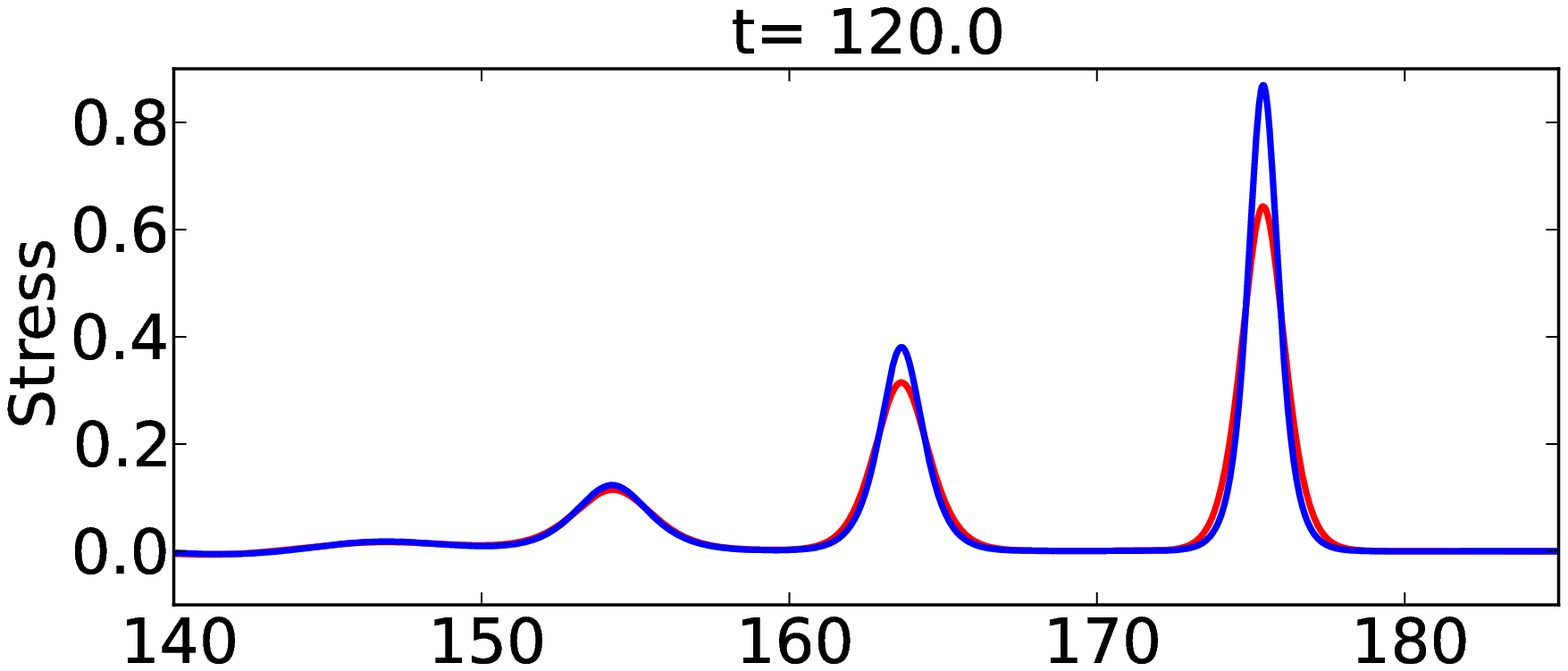}
\par\end{centering}
\caption{Diffracton solutions of \eqref{p-system} with the quadratic
nonlinearity \eqref{other nonlinearity}. 
We show a surface plot (left) of the stress at $t=120$ 
and slices (right) at the middle of material A (blue) and material B (red). 
\label{fig: other nonlinearity diffracton}}
\end{figure}

By the same token, diffractons may arise in the presence of quite
general nonlinearities, not just the exponential relation we have used.
As an example, Figure \ref{fig: other nonlinearity diffracton} shows
the solution of \eqref{p-system} obtained with the stress relation
\begin{align} \label{other nonlinearity}
 \sigma & = K(x)\epsilon+K(x)^2\epsilon^2,
\end{align}
with initial condition \eqref{gaussian}
where $K,\rho$ are given by \eqref{zmatched}.

\subsection{Wave propagation under reflection and diffraction}
\begin{figure}
\begin{centering}
\subfigure[Initial condition (close-up)\label{fig: other 2D diff-like waves, init condition}]{
	\includegraphics[scale=0.35]{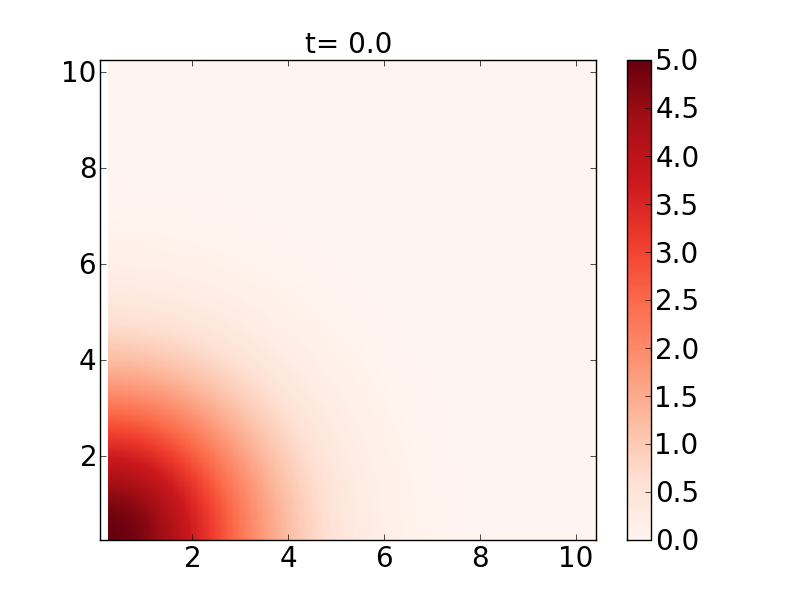}}
\subfigure[Mismatched impedance, matched sound speed\label{fig: other 2D diff-like waves, miss-matchZ_matchC}]{  
	\includegraphics[scale=0.35]{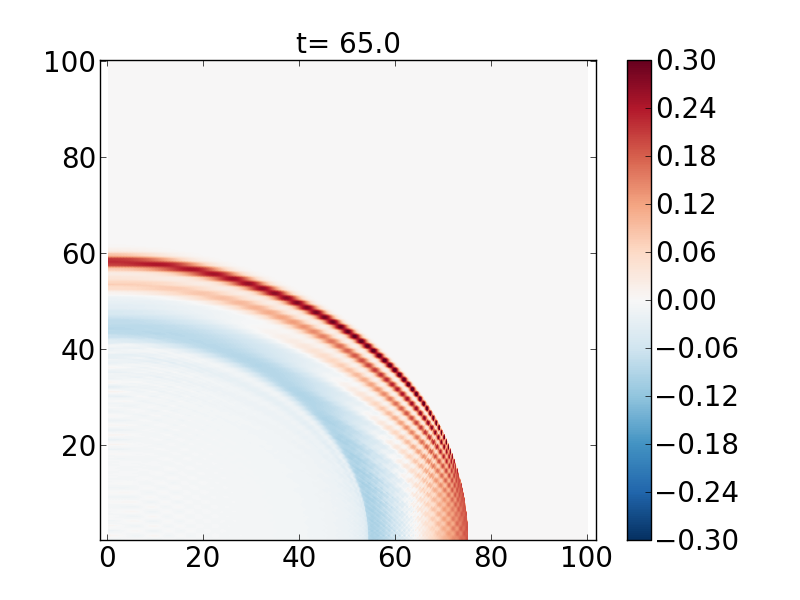}}

\subfigure[Matched impedance, mismatched sound speed\label{fig: other 2D diff-like waves, matchZ_miss-matchC}]{
  \includegraphics[scale=0.35]{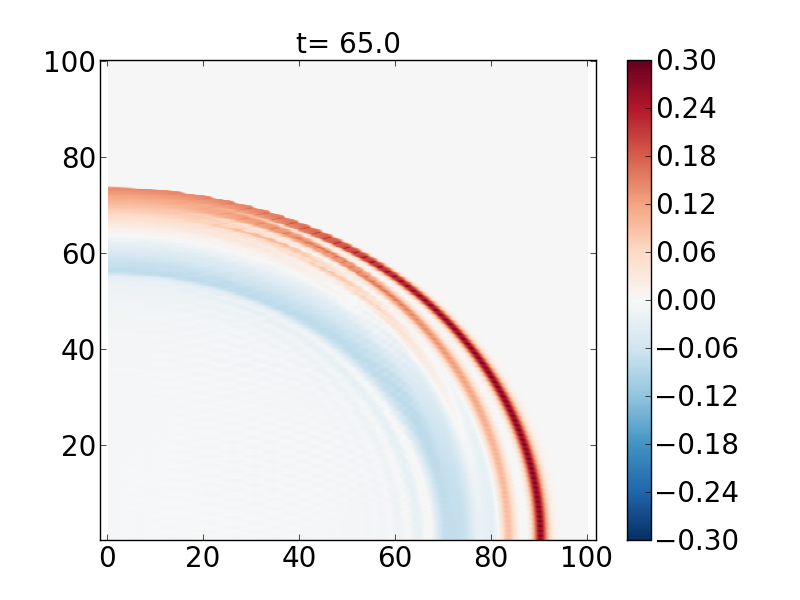}}
\subfigure[Mismatched impedance and sound speed\label{fig: other 2D diff-like waves, miss-matchZ_miss-matchC}]{  
  \includegraphics[scale=0.35]{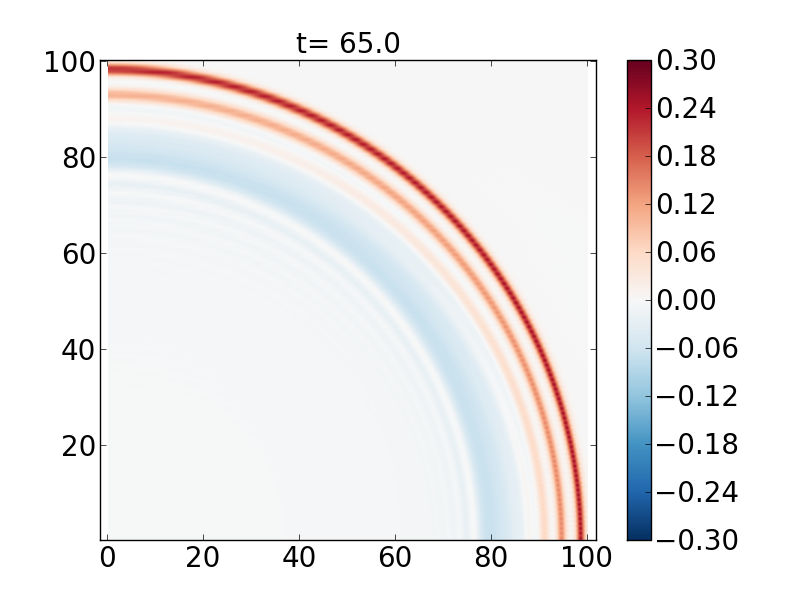}}
\par\end{centering}
\caption{Stress at $t=0$ and $t=65$ in different media. 
\label{fig: other 2D diff-like waves}}
\end{figure}

Waves that travel obliquely through a periodic medium like those considered here
undergo both reflection (if the impedance varies) and diffraction
(if the sound speed varies).  Thus in general they experience
two types of effective dispersion \cite{leveque2003,QK2013}.
Figure \ref{fig: other 2D diff-like waves} shows three experiments
demonstrating the possible scenarios.  For all three cases, the initial velocities are zero
and the initial stress (shown in Figure \ref{fig: other 2D diff-like waves, init condition}),
is
\begin{align}
\sigma(x,y,t=0) & = 5 \exp(-(x^2 + y^2)/10).
\end{align}
We show only the upper-right quadrant since the solution is symmetric.

In Figure \ref{fig: other 2D diff-like waves, miss-matchZ_matchC}, 
we take the material parameters \eqref{cmatched}, with the impedance 
mismatched and the sound speed matched. This generates dispersion by 
reflections. In this case, the solution develops a shock in the $x$-direction and 
solitary waves in the $y$-direction. 
In Figure \ref{fig: other 2D diff-like waves, matchZ_miss-matchC}, 
we use material parameters \eqref{zmatched} with the impedance matched and 
the sound speed mismatched. This introduces dispersion by diffraction. 
As a result, the wave develops a shock in the $y$-direction and solitary waves 
in the $x$-direction.
Finally, in Figure \ref{fig: other 2D diff-like waves, miss-matchZ_miss-matchC}, 
we consider a medium with $K_A=16$ and $\rho_A=K_B=\rho_B=1$
which leads to $Z_A=4$, $Z_B=1$, $c_A=4$ and $c_B=1$; i.e., the impedance and the 
sound speed are both mismatched. Effective dispersion due to reflections is introduced 
for waves traveling parallel to the $y$-axis, due to diffraction for waves traveling 
parallel to the $x$-axis and a combination of both if the wave travels 
in any other direction. A solitary wave develops that is almost cylindrically
symmetric.

\section{Conclusions and future work}
We have seen that typical solutions of the model \eqref{p-system} involve 
solitary wave trains, and that the effective dispersion responsible for these
waves is an effect of small-scale diffraction.  We think it would be useful
to better understand (from a physical point of view) how small-scale
diffraction (and reflection) lead to dispersive effects.
We are currently investigating the appearance of diffractive solitary
waves in other nonlinear wave models.

For large enough initial data, solutions of \eqref{p-system} may involve
shock waves.  These shock waves can turn into solitary waves after shedding
a sufficient amount of energy.  Investigation of shock wave formation and 
propagation in one- and two-dimensional periodic media is ongoing.

\appendix

\section{Numerical methods} \label{sec: numerical methods}
Solutions of the variable-coefficient system \eqref{p-system} in this work are
computed using PyClaw \cite{pyclaw-sisc}.
The algorithm used is referred to as SharpClaw and is
based on a WENO discretization in space and a SSP Runge-Kutta method in time
\cite{Ketcheson_Parsani_LeVeque_2011}.
For the Riemann solvers used and accuracy tests, see \cite{quezada2013}.

To solve the homogenized equations we use a pseudo-spectral spatial discretization from 
\cite{trefethen2000spectral} with a 4th order Runge-Kutta method in time. 

All code used to generate results in this work is available at
\url{https://github.com/ketch/diffractons_RR}.

\section{Derivation of homogenized equations} \label{appendix: homogenized equations}

The homogenized equations presented in Section \ref{sec:homog} are
derived through a nonlinear extension of the work in \cite{QK2013} applied to
system \eqref{2D psystem conservation form} with the 
constitutive relation \eqref{nonlinear_stress_relation}. Here we work through the
derivation; the reader is refered to \cite{QK2013} for even more detail.

Using \eqref{nonlinear_stress_relation}, 
we can write $\sigma_{\epsilon}=K\left(y\right)G\left(\sigma\right)$, 
where $G\left(\sigma\right)=\sigma+1$. 
Using the fact that $\sigma_t=\sigma_{\epsilon}\epsilon_t$, system 
\eqref{2D psystem conservation form} is transformed to:
\begin{subequations} \label{eq: Inverted system}
\begin{align}
 K^{-1}\left(y\right)\sigma_t-G\left(\sigma\right)\left(u_x+v_y\right) & = 0, \\
 \rho\left(y\right)u_t-\sigma_x & = 0,\\
 \rho\left(y\right)v_t-\sigma_y & = 0.
\end{align}
\end{subequations}

Next we introduce the small parameter $\delta$ explained in Section
\ref{sec:homog} and the fast scale $\hat{y}=\delta^{-1}y$.  By the chain rule, 
$\partial_y\mapsto\partial_y+\delta^{-1}\partial_{\hat{y}}$. 
Using the formal expansion
$\sigma\left(x,y,\hat{y},t\right)=\sum_{i=0}^\infty \delta^i \sigma_i \left(x,y,\hat{y},t\right)$
and similarly for $u$ and $v$, we get
\begin{subequations} \label{eq: expanded system}
\begin{align}
 K^{-1}\sum_{i=0}^\infty\delta^i\sigma_{i,t}-G\left(\sigma\right)\left(\sum_{i=0}^\infty\delta^i u_{i,x}+\sum_{i=0}^\infty\delta^i v_{i,y}+\delta^{-1}\sum_{i=0}^\infty\delta^i v_{i,\hat{y}}\right) & = 0, \\
 \rho\sum_{i=0}^\infty\delta^i u_{i,t}-\sum_{i=0}^\infty\delta^i \sigma_{i,x} & = 0, \\
 \rho\sum_{i=0}^\infty\delta^i v_{i,t}-\left(\sum_{i=0}^\infty\delta^i \sigma_{i,y}+\delta^{-1}\sum_{i=0}^\infty\delta^i \sigma_{i,\hat{y}}\right) & = 0,
\end{align}
\end{subequations}
where $(\cdot)_{i,x}$ denotes differentiation of $(\cdot)_i$ with respect to $x$. 
The function $G(\sigma)$ is expanded around $\sigma_0$ using Taylor series
as   
$G(\sigma)=G(\sigma_0)+\delta \sigma_1 + \delta^2 \sigma_2 + \dots$,
where we use the fact that $G'(\sigma)=1$ and all higher derivatives of $G$ vanish.

Next we equate terms of the same order in \eqref{eq: expanded system}. 
At each order we apply the averaging operator 
$\left\langle \cdot\right\rangle :=\int_{\mathcal C}\left(\cdot\right)d\hat{y}$ 
(where ${\mathcal C}$ denotes the unit cell of the medium)
to obtain the homogenized leading order system and corrections to it. 
At each order, we make an ansatz to obtain an 
expression for the non-homogenized solution.

\subsection{Derivation of the homogenized $\order(1)$ system}

Equating $\order \left(\delta^{-1}\right)$ terms in \eqref{eq: expanded system} 
and noting that $G\left(\sigma_{0}\right)\ne 0$ we conclude that
$v_0=v_0(x,y,t)=:\bar{v}_0(x,y,t)$ and $\sigma_0=\sigma_0(x,y,t)=:\bar{\sigma}_0(x,y,t)$, 
(the bar denotes variables that are independent of the fast scale $\hat{y}$).
We cannot conclude that $u_0$ is independent of the fast scale $\hat{y}$; 
indeed, we will soon see that it is not.
Now take the $\order(1)$ terms in \eqref{eq: expanded system} to get
\begin{subequations} \label{eq: Expanded O1 system}
\begin{align}
 K^{-1}\bar{\sigma}_{0,t}-G\left(\bar{\sigma}_0 \right)\left(u_{0,x}+\bar{v}_{0,y}+v_{1,\hat{y}}\right) & = 0, \\
 \rho u_{0,t}-\bar{\sigma}_{0,x} & = 0, \label{eq: Expanded O1 system, eqn2} \\
 \rho\bar{v}_{0,t}-\bar{\sigma}_{0,y}-\sigma_{1,\hat{y}} & = 0.
\end{align}
\end{subequations}
Divide the second equation by $\rho$ and apply the average operator 
$\left \langle \cdot \right \rangle$ to get: 
\begin{subequations} \label{eq: homog O1}
\begin{align}
 K_h^{-1} \bar{\sigma}_{0,t}-G\left(\bar{\sigma}_0\right)\left(\bar{u}_{0,x}+\bar{v}_{0,y}\right) & = 0, \\
 \rho_h\bar{u}_{0,t}-\bar{\sigma}_{0,x} & = 0, \label{eq: homog O1, eqn2}\\
 \rho_m\bar{v}_{0,t}-\bar{\sigma}_{0,y} & = 0,
\end{align}
\end{subequations}
where (based on periodicity) we have used
$\left\langle v_{1,\hat{y}}\right\rangle = \left\langle \sigma_{1,\hat{y}}\right\rangle =0$. 
Equation \eqref{eq: homog O1} is the homogenized leading order system. It has the same
form as \eqref{eq: Inverted system}, but with constant coefficients. 

From \eqref{eq: Expanded O1 system, eqn2} and \eqref{eq: homog O1, eqn2} 
one obtains the following relation between $u_0$ and $\bar{u}_0$:
\begin{equation}
 u_0=\frac{\rho_h}{\rho(\hat{y})}\bar{u}_0.\label{eq: fluctuating u0}
\end{equation}
This confirms that $u_0$ varies on the fast scale $\hat{y}$. Importantly, 
this shows that propagation in $x$ and the heterogeneity in $y$ are coupled
even at the macroscopic scale. 

\subsection{Derivation of the homogenized $\order(\delta)$ system}
In this section we first find an expression for the non-averaged
$\order\left(1\right)$ terms in \eqref{eq: Expanded O1 system}. 
To do so, we make the following ansatz: 
\begin{subequations} \label{eq: Ansatz for O1}
\begin{align}
 v_1 & = \bar{v}_1+A\left(\hat{y}\right)\bar{u}_{0,x}+B\left(\hat{y}\right)\bar{v}_{0,y}, \\
 \sigma_1 & = \bar{\sigma}_1+C\left(\hat{y}\right)\bar{\sigma}_{0,y}. \label{eq: Ansatz for O1, eqn2}
\end{align}
\end{subequations}
This ansatz is chosen in order to reduce system \eqref{eq: Expanded O1 system} 
to a system of ODEs. Substituting the ansatz \eqref{eq: Ansatz for O1}, 
the relation for $u_0$ \eqref{eq: fluctuating u0}, and the homogenized leading order
system \eqref{eq: homog O1} into the $\order \left(1\right)$ system 
\eqref{eq: Expanded O1 system} and equating the fast variable coefficients to zero, we get: 
\begin{subequations} \label{eq: ODEs for A, B and C}
\begin{align}
 A_{\hat{y}} & = K^{-1}K_h-\rho^{-1}\rho_h,\\
 B_{\hat{y}} & = K^{-1}K_h-1,\\
 C_{\hat{y}} & = \rho\rho_m^{-1}-1,
\end{align}
\end{subequations}
with the normalization conditions that 
$\left\langle A\right\rangle =\left\langle B\right\rangle =\left\langle C\right\rangle =0$. 
Note that $\left\langle A_{\hat{y}}\right\rangle =\left\langle B_{\hat{y}}\right\rangle =\left\langle C_{\hat{y}}\right\rangle =0$,
which implies that $A$, $B$ and $C$ are periodic. 

From \eqref{eq: expanded system} take terms of order $\order \left(\delta\right)$:
\begin{subequations} \label{eq: expanded Odelta system}
\begin{align}
 K^{-1}\sigma_{1,t}-G\left(\bar{\sigma}_0\right)\left(u_{1,x}+v_{1,y}+v_{2,\hat{y}}\right) 
 -\sigma_1\left(u_{0,x}+\bar{v}_{0,y}+v_{1,\hat{y}}\right) & = 0, \\
 \rho u_{1,t}-\sigma_{1,x} & = 0, \label{eq: expanded Odelta system, eqn2} \\
 \rho v_{1,t}-\sigma_{1,y}-\sigma_{2,\hat{y}} & = 0.
\end{align}
\end{subequations}

Plug the ansatz for $u_1$, $v_1$ and $\sigma_1$ from \eqref{eq: Ansatz for O1}
into \eqref{eq: expanded Odelta system} and take the average 
$\left\langle \cdot\right\rangle $ to get:
\begin{align*}
 K_h^{-1}\bar{\sigma}_{1,t}-G\left(\bar{\sigma}_0\right)\left(\bar{u}_{1,x}+\bar{v}_{1,y}\right) \nonumber 
 	-\bar{\sigma}_1\left(\bar{u}_{0,x}+\bar{v}_{0,y}\right) & = -\left\langle K^{-1}C\right\rangle \bar{\sigma}_{0,yt}
	+\left	\langle CB_{\hat{y}}\right\rangle \bar{\sigma}_{0,y}\bar{v}_{0,y} \nonumber\\
 	& \quad +\left(\rho_h\left\langle \rho^{-1}C\right\rangle +\left\langle CA_{\hat{y}}\right\rangle \right)\bar{\sigma}_{0,y}\bar{u}_{0,x},\\
  \rho_h\bar{u}_{1,t}-\bar{\sigma}_{1,x} & = \rho_h\left\langle \rho^{-1}C\right\rangle \bar{\sigma}_{0,xy},\\
  \rho_m\bar{v}_{1,t}-\bar{\sigma}_{1,y} & = -\left\langle \text{\ensuremath{\rho}A}\right\rangle \bar{u}_{0,xt}-\left\langle \rho B\right\rangle \bar{v}_{0,yt}.
\end{align*}

For many materials, including the layered and sinusoidal media considered in this work, 
it is true that 
$\left\langle K^{-1}C\right\rangle =\left\langle \rho^{-1}C\right\rangle =\left\langle CA_{\hat{y}}\right\rangle =\left\langle CB_{\hat{y}}\right\rangle =\left\langle \rho^{-1}C\right\rangle =\left\langle \rho A\right\rangle =\left\langle \rho B\right\rangle =0$. Therefore, we obtain:
\begin{subequations} \label{eq: homog Odelta}
\begin{align}
 K_h^{-1}\bar{\sigma}_{1,t}-G\left(\bar{\sigma}_0\right)\left(\bar{u}_{1,x}+\bar{v}_{1,y}\right)
 	-\bar{\sigma}_1\left(\bar{u}_{0,x}+\bar{v}_{0,y}\right) & = 0, \\
 \rho_h\bar{u}_{1,t}-\bar{\sigma}_{1,x} & = 0, \\
 \rho_m\bar{v}_{1,t}-\bar{\sigma}_{1,y} & = 0.
\end{align}
\end{subequations}

Since the boundary conditions are imposed in the leading order homogenized system, system \eqref{eq: homog Odelta} has zero boundary conditions
and no forcing terms; therefore, its solution vanishes:
\begin{equation*}
\bar{u}_1=\bar{v}_1=\bar{\sigma}=0.
\end{equation*}

\subsection{Derivation of the homogenized $\order(\delta^2)$ system}
First we make an ansatz for the non-averaged $\order(\delta)$ terms
$v_2$ and $\sigma_2$ in system \eqref{eq: expanded Odelta system}:
\begin{subequations} \label{eq: Ansatz for Odelta}
\begin{align}
v_2 & = \bar{v}_2+D(\hat{y})\bar{u}_{0,xy}+E(\hat{y})\bar{v}_{0,yy}, \\
 \sigma_2 & = \bar{\sigma}_2+F(\hat{y})\bar{\sigma}_{0,yy}+H(\hat{y})\bar{\sigma}_{0,xx}.
\end{align}
\end{subequations}
 
From \eqref{eq: expanded Odelta system, eqn2} we have $u_{1,t}=\rho^{-1}\sigma_{1,x}$. 
Then use the ansatz for $\sigma_1$ from \eqref{eq: Ansatz for O1, eqn2}
to get $u_{1,t}=\rho^{-1}C(\bar{\sigma}_{0,x})_y$ and the 
homogenized leading order equation \eqref{eq: homog O1, eqn2} 
to get $u_{1,t}=\rho^{-1}\rho_hC(\bar{u}_{0,y})_t$. 
Finally, we get an expression for the non-averaged solution $u_1$:
\begin{equation}
 u_1=\rho^{-1}\rho_hC\bar{u}_{0,y}.\label{eq: non-homogenized u1}
\end{equation}

Substitute the ansatz \eqref{eq: Ansatz for O1} for $v_1$ and $\sigma_1$,
the ansatz \eqref{eq: Ansatz for Odelta} for $v_2$ and $\sigma_2$,
the non-homogenized solution $u_1$ \eqref{eq: non-homogenized u1}, 
the leading order homogenized system \eqref{eq: homog O1}
and the ODEs \eqref{eq: ODEs for A, B and C} for $A$, $B$ and $C$
into \eqref{eq: expanded Odelta system} 
and set the fast variable coefficients to zero to get:
\begin{align*}
 D_{\hat{y}} & = K^{-1}K_hC-\rho^{-1}\rho_hC-A,\\
 E_{\hat{y}} & = K^{-1}K_hC-B,\\
 F_{\hat{y}} & = \rho\rho_m^{-1}B-C,\\
 H_{\hat{y}} & = \rho\rho_h^{-1} A,
\end{align*}
with the normalization condition
$\left\langle D\right\rangle =\left\langle E\right\rangle =\left\langle F\right\rangle =\left\langle H\right\rangle =0$
Again note that 
$\left\langle D_{\hat{y}}\right\rangle =\left\langle E_{\hat{y}}\right\rangle =\left\langle F_{\hat{y}}\right\rangle =\left\langle H_{\hat{y}}\right\rangle =0$, 
which implies $D$, $E$, $F$ and $H$ are periodic. 

From \eqref{eq: expanded system} take $\order(\delta^2)$ terms:
\begin{subequations} \label{eq: expanded Odelta2}
\begin{align}
 K^{-1}\sigma_{2,t}-G(\bar{\sigma}_0)(u_{2,x}+v_{2,y}+v_{3,\hat{y}})
 	-\sigma_1(u_{1,x}+v_{1,y}+v_{2,\hat{y}}) 
 	-\sigma_2(u_{0,x}+\bar{v}_{0,y}+v_{1,\hat{y}}) & = 0, \\
 \rho u_{2,t}-\sigma_{2,x} & = 0, \\
 \rho v_{2,t}-\sigma_{2,y}-\sigma_{3,\hat{y}} & = 0.
\end{align}
\end{subequations}

Plug the ansatz for $u_1$, $v_1$ and $\sigma_1$ from \eqref{eq: Ansatz for O1} 
and the ansatz for $u_2$, $v_2$ and $\sigma_2$ from \eqref{eq: Ansatz for Odelta}
into \eqref{eq: expanded Odelta2} and take the average
$\left\langle \cdot\right\rangle $ to get:
\begin{subequations} \label{eq: homog Odelta2}
\begin{align}
 K_h^{-1}\bar{\sigma}_{2,t}-G\left(\bar{u}_{2,x}+\bar{v}_{2,y}\right) \nonumber \\
 	-\bar{\sigma}_2\left(\bar{u}_{0,x}+\bar{v}_{0,y}\right) & = -K_h\left\langle K^{-1}F\right\rangle \left[G\left(\bar{u}_{0,xyy}+\bar{v}_{0,yyy}\right)+2\bar{\sigma}_{0,y}\left(\bar{u}_{0,xy}+\bar{v}_{0,yy}\right)\right] \nonumber \\
 	& \quad  -K_h\left\langle K^{-1}H\right\rangle \left[G\left(\bar{u}_{0,xxx}+\bar{v}_{0,xxy}\right)+2\bar{\sigma}_{0,x}\left(\bar{u}_{0,xx}+\bar{v}_{0,xy}\right)\right] \nonumber \\
 	& \quad +K_h\left\langle K^{-1}C^2\right\rangle \bar{\sigma}_{0,y}\left(\bar{u}_{0,xy}+\bar{v}_{0,yy}\right), \\
 \rho_h\bar{u}_{2,t}-\bar{\sigma}_{2,x} & = \rho_h\left\langle \rho^{-1}F\right\rangle \bar{\sigma}_{0,xyy}+\rho_h\left\langle \rho^{-1}H\right\rangle \bar{\sigma}_{0,xxx}, \\
 \rho_m\bar{v}_{2,t}-\bar{\sigma}_{2,y} & = -\rho_h^{-1}\left\langle \rho D\right\rangle \bar{\sigma}_{0,xxy}-\rho_m^{-1}\left\langle \rho E\right\rangle \bar{\sigma}_{0,yyy}.
\end{align}
\end{subequations}

\subsection{Combine homogenized leading order and corrections}

Once we have the homogenized leading order system and the homogenized corrections 
we combine them into a single system, using the relation
$\sigma:=\left \langle \sigma_0+\delta \sigma_1 +\dots \right \rangle$, and similarly 
for $u$ and $v$. Combining homogenized systems 
\eqref{eq: homog O1}, \eqref{eq: homog Odelta} and \eqref{eq: homog Odelta2}
we obtain:
\begin{align*}
 K_h^{-1}\sigma_t-\left(\sigma+1\right)\left(u_x+v_y\right) & = \delta^2\alpha_1\left[\left(\sigma+1\right)\left(u_{xyy}+v_{yyy}	\right)+2\sigma_y\left(u_{xy}+v_{yy}\right)\right]\\
 	& \quad  +\delta^2\alpha_2\left[\left(\sigma+1\right)\left(u_{xxx}+v_{xxy}\right)+2\sigma_x\left(u_{xx}+v_{xy}\right)\right]\\
 	& \quad +\delta^2\alpha_3\sigma_y\left(u_{xy}+v_{yy}\right),\\
 \rho_h u_t-\sigma_x & = \delta^2 \beta_1\sigma_{xyy}+\delta^2 \beta_2\sigma_{xxx},\\
 \rho_m v_t-\sigma_y & = \delta^2 \gamma_1\sigma_{yyy}+\delta^2 \gamma_2 \sigma_{xxy},
\end{align*}
where:
\begin{align*}
 \alpha_1 & = -K_h\left\langle K^{-1}F\right\rangle, &
 \alpha_2 & = -K_h\left\langle K^{-1}H\right\rangle, &
 \alpha_3 & = K_h\left\langle K^{-1}C^2\right\rangle, \\
 \beta_1 & = \rho_h\left\langle \rho^{-1}F\right\rangle, &
 \beta_2 & = \rho_h\left\langle \rho^{-1}H\right\rangle \\
 \gamma_1 & = -\rho_m^{-1}\left\langle \rho E\right\rangle, & \gamma_2 & = -\rho_h^{-1}\left\langle \rho D\right\rangle.
\end{align*}
Formulas for these coefficients in the case of a piecewise constant 
medium are given by \eqref{layered_medium_coefficients}.

\bibliographystyle{plain}
\bibliography{bibs}
\end{document}